%% file: almapublicationstatistics.tex
\documentclass[twocolumn,tighten,bookmarks=true,pdfnewwindow=true,colorlinks=true,linkcolor=xlinkcolor,citecolor=xlinkcolor,filecolor=xlinkcolor,urlcolor=xlinkcolor,final=true]{openjournal}
\usepackage{graphicx}
\usepackage{lastpage}

\usepackage[usenames,dvipsnames]{xcolor}
\definecolor{xlinkcolor}{cmyk}{1,1,0,0}

\makeatletter
\renewcommand\@makecaption[2]{%
  \par
  \vskip\abovecaptionskip
  \begingroup
    \footnotesize\rmfamily
    \begingroup
      \samepage
      \flushing
      \let\footnote\@footnotemark@gobble
      \ifnum\pdfstrcmp{\@captype}{table}=0
        \@make@capt@title{\textsc{Table \thetable}}{#2}%
      \else
        \ifnum\pdfstrcmp{\@captype}{figure}=0
          \@make@capt@title{\textsc{Figure \thefigure}}{#2}%
        \else
          \@make@capt@title{#1}{#2}%
        \fi
      \fi\par
    \endgroup
  \endgroup
  \vskip\belowcaptionskip
}
\makeatother

\usepackage{etoolbox}
\AtBeginEnvironment{tabular}{\renewcommand{\arraystretch}{1.5}}
\AtBeginEnvironment{tabularx}{\renewcommand{\arraystretch}{1.5}}
\usepackage{orcidlink}
\usepackage[hang,flushmargin]{footmisc}
\usepackage{url}

\begin{document}
   \title{ALMA publication statistics\vspace{-4.25em}}   
   \author{Felix Stoehr$^{1,\star}$, Mar\'ia\ D\'iaz Trigo$^1$\orcidlink{0009-0006-7491-9396}, Evanthia~Hatziminaoglou$^{1,2,3}$\orcidlink{0000-0001-7796-4279}, Uta Grothkopf$^1$\orcidlink{0000-0001-6830-0702}, Silvia Meakins$^1$\orcidlink{0000-0002-4911-133X}, Leslie Kiefer$^4$\orcidlink{0000-0002-1774-6242}, Lance~Utley$^5$\orcidlink{0009-0007-7866-8800}, Mika~Konuma$^6$\orcidlink{0000-0003-0917-9636}, Eelco van Kampen$^1$\orcidlink{0000-0002-6327-5154}, Gergö Popping$^1$\orcidlink{0000-0003-1151-4659}, Enrique~Macias$^1$\orcidlink{0000-0003-1283-6262}, Martin Zwaan$^1$\orcidlink{0000-0003-0101-1804}}

   \affiliation{$^1$ European Southern Observatory, Karl-Schwarzschild-Str. 2, D-85487 Garching, Germany}
   \affiliation{$^2$ Instituto de Astrof\'{i}sica de Canarias, 38205 La Laguna, Tenerife, Spain}
   \affiliation{$^3$ Departamento de Astrof\'{i}sica, Universidad de La Laguna, 38206 La Laguna, Tenerife, Spain}
   \affiliation{$^4$ ALMA Santiago Central Offices, Alonso de Córdova 3107, Vitacura, Santiago, Chile}
   \affiliation{$^5$ National Radio Astronomy Observatory, 520 Edgemont Road. Charlottesville, USA}
   \affiliation{$^6$ National Astronomical Observatory of Japan, 2-21-1 Osawa, Mitaka, Tokyo 181-8588, Japan}
   
\thanks{$^{\star}$E-mail: fstoehr@eso.org}

\begin{abstract}
The success of an astronomical facility is measured by its scientific impact. A principal metric for this impact is the ensemble of peer-reviewed publications based on the observational data obtained by the facility. We present a comprehensive study of the statistics of the 4,190 refereed publications of the Atacama Large Millimeter/Submillimeter Array (ALMA) in the period from 2012 to 2024. The publications have received 169,985 citations and are based on 2,670 ALMA projects totalling 19,265 hours of 12-m-array-equivalent observing time. Our study analyses publication statistics related to various aspects, e.g. science categories, geographical distribution, archival research, time to publication, publication fraction, and citations. We also look into the community and compare ALMA with other facilities. We find that ALMA is a high-impact observatory with an average of 41 citations per publication, $\sim$70\% of observed projects published, $\sim$40\% of publications making use of archival data in 2024, more than 9,400 unique authors, and a publication evolution following that of HST and VLT. Currently, the impact factor for ALMA publications is larger than that of all other major astronomical facilities. ALMA also plays a pivotal role in very long baseline interferometry (VLBI), substantially contributing to landmark achievements such as capturing the first image of a black hole shadow.
\end{abstract}

\keywords{ALMA -- publication -- statistics -- citations -- impact -- archive}

\maketitle
\tableofcontents

\section{Introduction}
Knowledge discovery in the scientific process is communicated through scientific articles. In astronomy, scientists publish their findings in journals, where peer review ensures high standards of quality. The number of publications based on observations from an observatory is generally used as a measure of the scientific progress enabled by, and the impact of, that facility. Major observatories provision libraries whose staff follow refereed journals, carefully evaluating whether or not a given article makes actual use of the facility's observations, and the results are routinely published and made available to funding agencies. Primarily, however, the goal of curating publication information and analysing their statistics is to gain insights, enabling further optimisation of the strategy of the observatory and science operations 
\citep[e.g.][]{2010PASP..122..808A_Apai, 1110.1212_Rots, 2014SPIE.9149E..0A_C_Crabtree, 2014AN....335..210N_Ness, 2015Msngr.162...30S_Stoehr, 2016SPIE.9910E..03S_Sterzik, 2017msngr.169...11L_Leibundgut, 2025AN....34670014_Ness}.

Although the number of publications is the most widely used measure of scientific productivity, it is understood that this number is affected by several factors. These include, for example, the size of the astronomical community - itself varying with time - capable of working with the facility's data, the usability of the data to address astronomical topics of interest at a particular time, the scientific field \citep[e.g.][]{2012PLoSO...746428P_Pepe}, or the ease with which data from a facility can be used to extract meaningful new science.

Beyond counting the number of publications, second order statistics are thus routinely used to characterise the scientific output of an observatory, such as the number of citations the publications of a facility receive - again, strongly varying with time - or the impact factor, where the number of citations received is normalised by the number of publications.

Although publication statistics are important for observatories, it is also widely recognised that they can only provide a proxy measurement of the actual scientific impact of a facility. Furthermore, interpretation of the numbers using additional context is necessary but may also be subjective. Finally, as much as the statistics are relevant for optimising science operations, the effect of a single particular measure taken by the observatory is generally impossible to disentangle from the effects occurring in the years between data availability and the appearance of corresponding publications.

In this work, we present statistics of the refereed publications using data from the Atacama Large Millimeter/submillimeter Array (ALMA), covering the period from the first ALMA publication in 2012 up to and including 2024. After a brief summary of ALMA (Section~\ref{section:ALMA}), we describe the methodology used (Section~\ref{section:methodology}) and then present the individual statistics (Section~\ref{section:statistics}) before concluding (Section~\ref{section:conclusions}).

\section{ALMA}
\label{section:ALMA}
ALMA is a millimetre/submillimetre radio-interferometer located in the Atacama Desert in Chile at an altitude of 5,000 metres above sea level. It consists of three antenna arrays: one array of 50 antennas with a 12-m diameter, one array of 12 antennas with a 7-m diameter, as well as one array of four additional antennas with a 12-m diameter for Single Dish (or Total Power, hereafter TP) measurements. The latter two arrays are also referred to as the ALMA Compact Array (ACA). Observations with these arrays can be carried out simultaneously and independently of each other. The arrays cover complementary angular scales on the sky and - depending on the science case and the angular extension of the source - observations with one, two, or three arrays are carried out for a source.

ALMA issues yearly calls for proposals offering a total number of hours of observing time that ramped up from 500 hours in Cycle 0 to the nominal value of 4,300 hours since Cycle 7\footnote{Offered hours in the calls for proposals: Cycle 0: 500-700h, Cycle 1: 800h, Cycle 2: 2000h, Cycle 3: 2,100h, Cycle 4: 3,000h, Cycle 5: 4,000h, Cycle 6: 4,000h, Cycles 7-11: 4,300h}. After a highly competitive process, the observing time is awarded to typically 200-300 ALMA projects per year. These consist of projects scheduled at high priority (Grades A and B), out of which Grade~A projects are carried over to the next cycle if they are not completed. In addition, 200-300 filler projects (Grade~C) are selected, which are executed if the observing conditions are such that no A or B project can be observed. Projects are selected based entirely on scientific merit while, at the same time, following the distribution of the time-allocation shares of the three ALMA partners: East Asia 22.5\%, Europe 33.75\%, and North America 33.75\%, as well as Chile with 10\%. ALMA is open to receiving proposals from astronomers worldwide (Open Skies). The observing time of projects unaffiliated with an ALMA partner (see Section \ref{subsection:countries}) is attributed to the regions proportionally to their share, up to a total of 5\% of the cycle's observation time, and to North America beyond that threshold. Up to 5\% of the observing time in a Cycle can be allocated by the ALMA Director through Director’s Discretionary Time (DDT).

The observing capabilities of ALMA have been continuously augmented, from 16 antennas and three receiver bands at the start of ALMA operations in 2011 up to nine receiver bands\footnote{The tenth receiver band, Band 2, will be offered as of 2026.} and the full set of 66 antennas in 2024. In Cycle 4 (2016) Large Programs (LP) were introduced, allowing Principal Investigators (PI) to request projects with more than 50 hours of observing time. Also, observing modes like mosaic, solar, Very Long Baseline Interferometry (VLBI), ACA stand-alone, and full polarisation  observations were added gradually, typically preceded by the release of Science Verification (SV)\footnote{\url{https://almascience.org/alma-data/science-verification}} data demonstrating the new capability (see also Subsection~\ref{subsection:scienceverification}). 

ALMA was designed to attract scientists beyond those who are experts in interferometry, millimetre/submillimetre wavelengths, or any given scientific field. Today, ALMA is used by a large community of researchers (see Section~\ref{subsection:community}) thanks to the availability of high-quality data and data products delivered by the observatory, the ease of data discovery and data access, and the extensive user support \citep[e.g.][]{2015Msngr.162...24_Hatziminaoglou}. The observations are protected by a 12-month proprietary period before they are publicly accessible. 
The proprietary period for DDT projects is six months for the data considered in this work.\footnote{In 2025, the observatory removed the proprietary period for DDT projects unless an exception is granted.}

The smallest unit of ALMA observing instructions is called a Scheduling Block (SB). A SB contains calibration observations as well as science observations for one or more science targets. The data, observed by executing the SB one or more times until the required sensitivity is reached, are called a Member Observing Unit Set (Member OUS). The Member OUS is the smallest unit of data that is processed by the ALMA pipeline and delivered to the PI.

\section{Methodology}
\label{section:methodology}
Library staff at the European Southern Observatory (ESO), National Radio Astronomy Observatory (NRAO), National Astronomical Observatory of Japan (NAOJ) and Joint ALMA Observatory (JAO) are analysing refereed publications appearing in a large number of journals with the help of dedicated software \citep[e.g.][]{2005Msngr.119...50D_Delmotte, 2010ASPC..433...81E_Erdman, 2015ASPC..492...63G_Grothkopf} and are curating the bibliography of their respective facilities, including ALMA. For this work, we use the state of the bibliography database as of the 1st of July 2025.

\begin{figure}[th]
  \resizebox{\hsize}{!}{\includegraphics{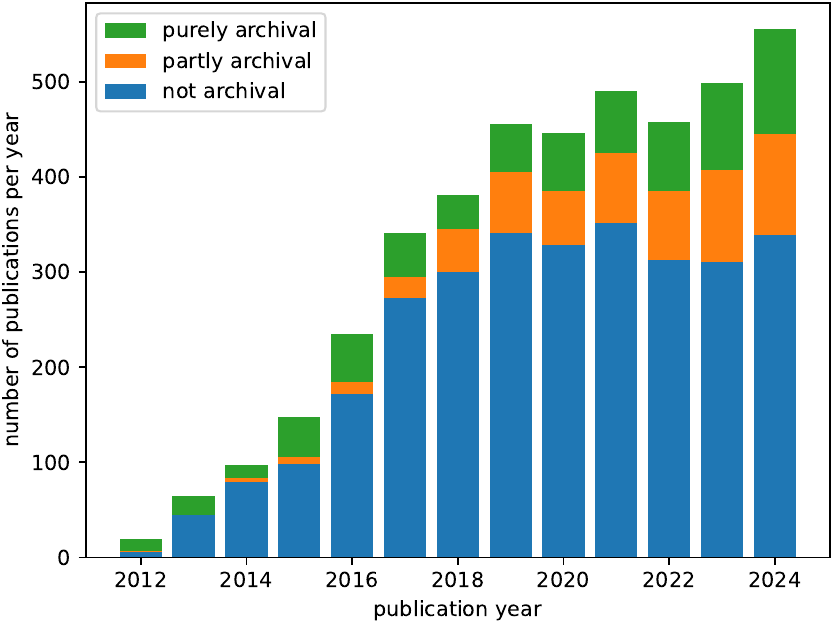}}
  \caption{Stacked histogram of the evolution of the number of ALMA publications per year. In 2024, for the first time, more than 500 publications were recorded in one year. Whereas the number of publications making only use of PI data (blue) has been roughly constant since 2019, the number of publications using PI data and archival data together (orange) and the number of publications making pure use of archival data (green) are both growing.}
  \label{figure:evolutiongeneral}
\end{figure}

The ALMA bibliography includes publications that make direct use of ALMA data. In line with practice at e.g. Hubble Space Telescope (HST), James Webb Space Telescope (JWST), {\it Spitzer} and Very Large Telescope (VLT)/VLT Interferometer (VLTI) \citep{2024OJAp....7E..85O}, publications that only cite previous results, discuss instrumentation or software, mention projects, use data as examples, or show images without scientific analysis are excluded\footnote{\url{https://eso.org/sci/libraries/telbib_methodology.html}}. Data of a project used in a publication are classified as `archival' in that publication if there is no overlap between the list of authors of the publication and the list of the PIs and Co-Investigators (CoIs) of the project. This definition is conservative compared to some other facilities where e.g. data are declared archival as soon as the PI is not part of the author list of the publication \citep{2024arXiv240212818D_DeMarchi}.

ALMA is the first astronomical observatory to have required by policy that authors using ALMA data {\it must} acknowledge the project codes of the data they have used in their publication\footnote{\url{https://almascience.eso.org/documents-and-tools/latest/alma-user-policies}}. ALMA monitors the arXiv\footnote{\url{https://arxiv.org/archive/astro-ph}} pre-print service continuously and politely reminds the authors via email of this obligation should they have forgotten to add the statement into their publication. This practice - which now also has been established for the JWST (Josh Peek, private communication) - ensures a high quality and completeness of the project code information needed for data curation.

Also, ALMA runs an anonymous survey on data that have not been published - first two, since 2023 three years after the data have been made available to the PI - to identify areas where observatory operations can be improved \citep{2016arXiv161109625S_Stoehr}. It happens that in this process the PI indicates a publication that was overlooked because e.g. a typo had been present in the project code or because the journal was not yet monitored, furthering the completeness of the ALMA bibliography. Overall, we estimate a completeness fraction of more than 95\% for the ALMA corpus.

While the granularity of the use of ALMA data to be acknowledged is the project level, ALMA staff extend the information by reading the publications referencing ALMA data from more than one band or array and curating the bands and arrays that were actually {\it used} into the bibliography database (see also Subsection~\ref{subsection:receiverband}).

Several statistics provided below require a measure of observing time. We follow ALMA's definition counting the execution time of the SB including overheads. The total observing time of a project making use of several arrays is computed by normalising the observing times of the 7-m and TP arrays to the observing times of the 12-m array by the total antenna surface of the arrays. One hour on the 7-m and TP arrays corresponds to roughly 0.077 hours and 0.068 hours on the 12-m array, respectively. The observing time in this work is expressed in `12-m-array-equivalent' hours.

We obtain the number of citations for each publication from the Astrophysics Data System (ADS)\footnote{\url{https://scixplorer.org}} \citep{2016ASPC..512..473A_Accomazzi} which we also store on a weekly basis in our database to construct the full citation evolution history for each paper. 

\section{Statistics}
\label{section:statistics}
\subsection{General}
\label{section:general}
Figure~\ref{figure:evolutiongeneral} shows the evolution of the total number of refereed publications making use of ALMA data. A phase of rapid increase up to 2019 is since then followed by a continued but slower increase, partly likely influenced by the pandemic. In 2024, the yearly number of publications based on ALMA data reached 555 and thus an all-time high. While the number of publications making only use of PI data (blue) has been roughly constant since 2019, the growth is driven by publications making use of PI data and of archival data together (orange), as well as by publications made purely from archival data (green). In 2024, the fraction of publications making use of any archival data was 39\%. For a more detailed discussion of archival publications see Subsection~\ref{subsection:archivalpublications}.

\subsection{Science categories}
\label{subsection:sciencecategories}
\begin{table*}[t]
    \centering
    \centering
    \begin{tabular}{l r r r r r r r r r}
\hline\hline
\input{table_01.tex}

    \end{tabular}
    \caption{Summary of the PI projects that have been observed by ALMA, the corresponding publications and citations per scientific category. We use our the science category definition from the ASA which is stable in time and is based on the scientific keywords. We distribute the citations and the (fractional) publication over science categories of the projects weighted by the 12-m-equivalent observing time and split over the categories in case a project has keywords from different categories. For the column citations/project we normalise by the published projects. For the last two columns - publications and citations per hour -, the total fractional 12-m-array equivalent observing time of the observed projects of each science category was used. For comparison, a line has been added below the table that includes summary values for all ALMA projects, i.e. including the ALMA observatory projects SV, E and CAL (see Section~\ref{subsection:observingmode}).}
    \label{table:sciencecategories}
\end{table*}
When submitting a proposal, PIs indicate which scientific category the proposal corresponds to as well as one or two scientific keywords. As the names of the typically five scientific categories offered to the PIs in the ALMA Observing Tool (OT)\footnote{\url{https://almascience.org/proposing/observing-tool}} have changed over time, we follow the methodology of the ALMA Science Archive (ASA) and start from the scientific keywords provided by the PIs, which we then map to the nine science categories shown in Table~\ref{table:sciencecategories}. These categories are independent of the observing cycle. Table~\ref{table:sciencecategories} provides a summary of statistics for those scientific categories.

Each PI project that has received any data, for all three proposal grades A, B and C, is counted in the first column. Projects having a single science keyword or projects where both science keywords point to the same science category contribute to that category. For projects that have two scientific keywords pointing to two different categories, we assign half the project to each of those categories so that the total value indeed reflects the actual total number of observed PI projects. The second column lists the projects that have been used in at least one publication - be it by the PI group or through archival research - again split onto the science categories in the same way. We exclude projects from being marked as published if they appear only in publications that use a large fraction of all ALMA calibrator data \citep[i.e. the ALMACAL series starting with][]{2016ApJ...822...36O_Oteo}. For the third column, we compute the fractional publication weighted by the 12-m-array-equivalent observing time of each project\footnote{only taking into account the bands that have actually been used (see Section~\ref{subsection:receiverband})} and then distribute it for each project over the science keywords. The same methodology is used to distribute the cumulative citations the publication has received (independent of the age of the publication) over the science categories. The fifth and sixth columns show these citations normalised by the published projects and the number of publications, respectively. The citation fraction shows the values of the citation column normalised, and the last two columns show the number of publications and the number of citations normalised by the 12-m-array-equivalent observing time of the observed projects again distributed over the categories using the science keywords.

Finally, the line at the bottom shows the same total summary as just above, but this time for all ALMA projects, i.e. including the observatory projects SV, E and CAL (see Sections~\ref{subsection:scienceverification} and \ref{subsection:observingmode}). In total 4,190 refereed articles making use of ALMA data have been published up to the end of 2024 and they have received 169,985 citations.

In terms of the number of publications, `ISM and star formation' dominates with 1162 and thus over 50\% more publications than `Disks and planet formation' as the second category. The largest fraction of citations is accumulated by the science categories `Active galaxies', `ISM and star formation', `Disks and planet formation' and `Galaxy evolution'. 

We find that the number of observed projects has a large spread between the categories with a factor of 36 between `ISM and star formation' and the `Sun'. Per invested hour of observing time, the values are far more homogeneous with a spread of only 2.4. Outside of the two categories `Disks and planet formation' and `Active galaxies' with the largest value of citations per invested hour, the remaining categories show remarkably comparable values. It may be relevant for the proposal review process, that the science category with the largest number of citations per hour of observing time is only the fourth-largest category in terms of observed projects.

\subsection{Receiver bands}
\label{subsection:receiverband}
\begin{figure}[t]
  \resizebox{\hsize}{!}{\includegraphics{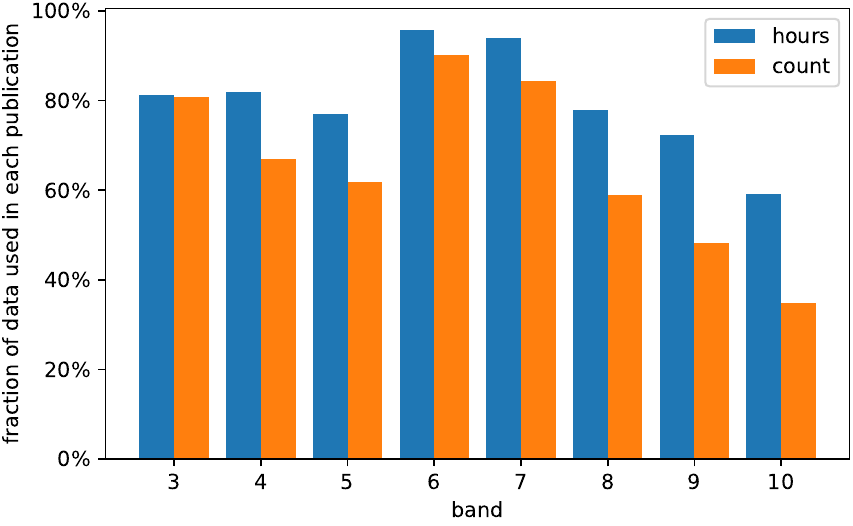}}
  \caption{Bands actually {\it used} in a given publication as opposed to the data of a given band that were part of the observations in the projects that the publication utilized. The statistics are given for the 12-m-array-equivalent hours of the observations of a given band used in a publication (blue) as well as just counting the bands (orange). On average, if one or more projects have been included in a publication that have observations in Band 10, then in about 35\% (but 59\% of the hours) of these publications Band 10 was actually used for the science. }
  \label{figure:bandsused_vs_bandsobserved}
\end{figure}
The publications up to 2024 are based on data from the eight ALMA receiver bands, Band 3 through Band 10\footnote{Note that Band 1 was only offered in 2024, and therefore no corresponding publications are included in the set of publications analysed here.}. At any given time, each array can only observe in a single receiver band. ALMA projects can contain data from several receiver bands, and also a publication can make use of data from several projects. To understand which bands were actually {\it used} in a publication as opposed to the bands contained in the {\it projects} the publication was based on, the text of the publications is analysed. ALMA staff members read publications that use data from projects involving more than one band or array and record the information (see Section~\ref{section:methodology}). This information is presented in Figure~\ref{figure:bandsused_vs_bandsobserved}.
 
We find, for example, that 62\% of the publications based on projects that contain some Band 5 data actually make use of that Band 5 data. Similarly, in terms of observing hours, on average, 77\% of the Band 5 data contained in projects that the publications are based on, are actually used. On the other hand, in nearly all cases where Band 6 or Band 7 data were part of the projects on which a publication was based, those Band 6 or Band 7 data were also truly used. The fractions of usage of the other bands are roughly consistent with each other, except for Band 10, where in about 35\% of the cases the data were actually used, and about 59\% of the hours were used. 

\begin{table}[b]
    \centering
    \begin{tabular}{l r r r r r}
      \hline\hline
       \input{table_02.tex}
    \end{tabular}
    \caption{Distribution of the number of refereed publications actually making use of a given ALMA Receiver Band. The second column shows the corresponding fraction of all publications. As in a publication data from several receiver bands can be used, the sum of the number of publications is larger than the total number of publications and the fractions add up to 150\%. The third column shows the fractions renormalised to 100\%, the fourth column shows the total fraction of citations produced by a given receiver band and the last column shows the number of citations a particular band received normalised by the 12-m-array equivalent number of corresponding ALMA observations. This table contains all publications including publications making use of observatory projects.}
    \label{table:receiverbands}
\end{table}

\begin{figure}[t]
  \resizebox{\hsize}{!}{\includegraphics{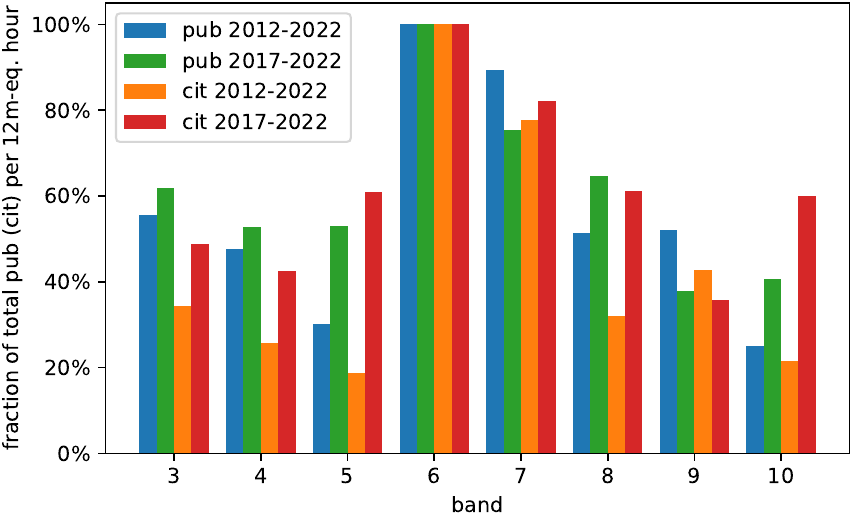}}
  \caption{The scientific productivity - here measured in terms of the number of publications and the number of citations - per invested hour of observing time as a function of the receiver band. For each publication and its corresponding citations, only the bands that were actually used in a given publication have been counted (see Figure~\ref{figure:bandsused_vs_bandsobserved}) and the fractional publication (or citations) have been distributed over the receiver bands and projects weighted by the 12-m-array-equivalent observing hours. Four measurements are shown: in blue the number of publications, taking into account projects from Cycle 1 to Cycle 11, i.e. excluding Cycle 0 as well as observatory projects. In orange the same measurement but counting citations instead of publications. As not all receiver bands were available from Cycle 1 onward, there is a bias in those numbers. To reduce that bias, we also show in green and red the equivalent numbers but now restricting to cycles where all receiver bands were offered. For better comparison, all four measurements were normalised to their largest value, i.e. Band 6 in all cases.}
  \label{figure:publications_citations_per_band}
\end{figure}

The lower numbers for Band 10 likely reflect the combined impact of technical challenges in calibration and data quality at very high frequencies, the relatively late coming of age of the corresponding analysis tools, and the smaller and more exploratory nature of the first Band 10 projects. As time goes by, this gap is expected to narrow.

In Table~\ref{table:receiverbands} we present publication and citation statistics per ALMA band that were actually {\it used} in each publication. The first and second columns show the number and fraction of publications in which the band was used, respectively. As several bands can be used in a publication, these numbers are larger than the total number of publications and 100\%, respectively. Normalising the second column gives the third column. The last two columns show the fraction of citations that the receiver band obtained as well as the citations per invested hour of observing time. To compute these values, the citations received for a given publication have been distributed over the receiver bands of the data in that publication proportionally to the 12-m-array-equivalent observing hours of the data that were actually used in the publication.

Note that there are three large biases to keep in mind: 1) not all bands have received the same amount of observing time, 2) not all bands have been available on the telescope for the same amount of time, and 3) the table does include observatory projects, which typically receive a large number of citations for a short observing time. These biases are addressed in Figure~\ref{figure:publications_citations_per_band} showing the scientific productivity - here measured in terms of the number of publications and the number of citations - of the different receiver bands, normalised by the 12-m-array-equivalent observing hours distributed over the projects used in the publication\footnote{Note that we have no information about how much each Band inside a given publication - making use of several bands - does contribute to the scientific result and consequently to the citations, and thus assume the weighting by observing-time.}. For better clarity, all values have been individually normalised to Band 6, which has the largest productivity in all cases. The figure only shows PI data to remove any bias that might stem from SV projects. Two periods of time are shown, once all projects from cycles 2012-2022 and once projects from 2017-2022 and thus only the the projects for the years in which all ALMA bands were online. In both cases we stop in 2022, to account for the fact that it takes time to accumulate citations for a publication. The choice of two years for the cut-off, i.e. usage of projects until 2022 and publications until 2024, is motivated by the median value of the time between data delivery and publication (see Figure~\ref{figure:publicationdelay}). 

\begin{table}[t]
    \centering
    \begin{tabular}{l r r r r}
        \hline\hline
        \input{table_03.tex}
    \end{tabular}
    \caption{Distribution of refereed publications from 2012 to 2023. Publications in journals with less than 1\% total fraction each have been combined into {\it other}. Shown are the fraction of the total number of publications, the total number of publications as well as the number of citations for each journal.}
    \label{table:journals}
\end{table}

Concentrating on the publications of projects between 2017 and 2022 where all bands were online, the results between citations and publications are remarkably consistent. While Bands 6 and 7 clearly stand out and deliver the highest fractions of citations and fractional publications, all other bands show comparable numbers of about 40\% to 60\% of the productivity of Band 6. Note that as only 18 publications have made use of Band 10 data, the values are affected by small number statistics.

There is an intriguing difference between the values for Band 9 and Band 10 between the Table~\ref{table:receiverbands} and Figure~\ref{figure:publications_citations_per_band} which is worthwhile discussing. The value of citations per hour for Band 9 in the table is 5.8 and thus the second highest after Band 6, whereas in the figure the values for Band 9 for projects 2017-2022 are the lowest of all bands. The reason for this difference lies in the exclusion of projects from Cycle 0 and the SV data. The Cycle 0 data and the SV data are extremely well cited. For most bands, Cycle 0 and SV/E/CAL only corresponds to a very small fraction of the total time of all data observed. Except for Band 9 where about 16\% of all observing time is in such projects. For Band 10, which came online only later\footnote{Bands 3, 6, 7, 9 were offered in 2011, Bands 4, 8 in 2013, Band 10 in 2015, Band 5 in 2017}, the bias 2) mentioned above is very strong and responsible for the difference in values. The most relevant numbers comparing bands for observations allocated in future proposal review processes are therefore the values for the projects 2017-2022 in Figure~\ref{figure:publications_citations_per_band}.

\subsection{Journals}
\label{subsection:journals}
Table~\ref{table:journals} lists the overall number of publications per journal for all journals with a share of 1\% or larger. The vast majority (88\%) of all ALMA results are published in just four journals: ApJ (35\%), A\&A (25\%), MNRAS (17\%) and ApJL (11\%).

\begin{figure}[t]
  \resizebox{\hsize}{!}{\includegraphics{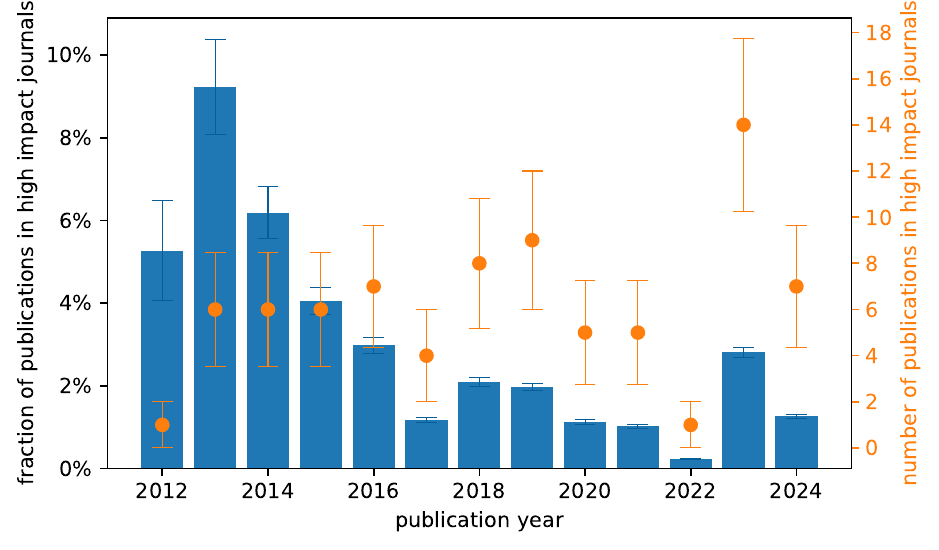}}
  \caption{Evolution of the fraction (blue) and number (orange) of ALMA publications in the high-impact journals {\it Nature} and {\it Science}.}
  \label{figure:evolution_journals}
\end{figure}

Of particular interest to observatories is the number and fraction of publications in the high-impact journals {\it Nature} and {\it Science} (excluding derivatives like {\it Nature Astronomy} or {\it Science Advances}), serving as an indicator of major new science of general interest. Figure~\ref{figure:evolution_journals} shows the evolution of the fraction of such publications (blue) as well as their number (orange). At the start of observations, opening a new window to the Universe, nearly 9\% of all ALMA publications were published in the two high-impact journals, dropping over the years to roughly 1\%-2\% when the observatory had matured. 

At the same time, with the growing total number of publications, while the fraction dropped, the average number of publications in high-impact journals is 6.1 per year with no sign of decline.

\subsection{Countries}
\label{subsection:countries}
First authors from institutes in 52 countries and six major international organisations - ALMA, ESO, ESA, SKA, IRAM, EHT which are counted separately - have been publishing using ALMA data. This is more than double the number of `ALMA countries', defined here as the countries associated with the three ALMA partners ESO (Austria, Belgium, Czechia, Denmark, Finland, France, Germany, Ireland (joined ESO in 2018), Italy, The Netherlands, Poland (joined ESO in 2015), Portugal, Spain,  Sweden, Switzerland, United Kingdom), NAOJ (Japan, Taiwan, Republic of Korea (joined in 2014)) and NRAO (USA, Canada). The host country Chile is counted separately.

\begin{figure}[t]
  \resizebox{\hsize}{!}{\includegraphics{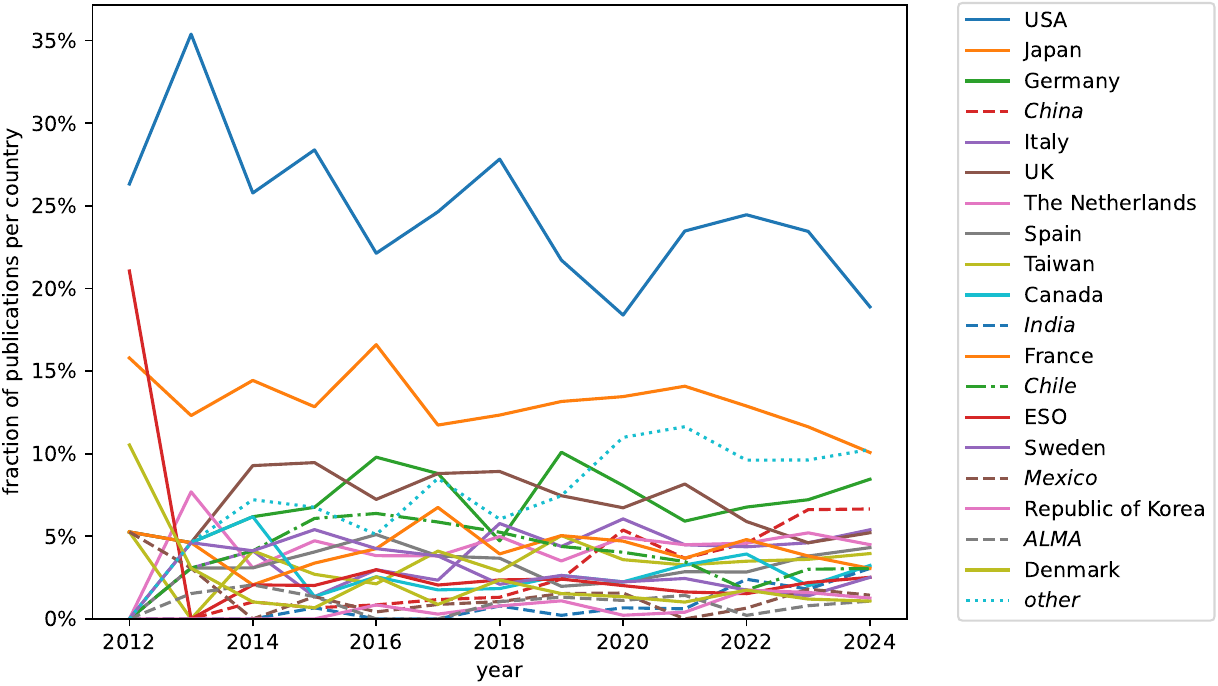}}
  \caption{The fraction of publications per country is shown as a function of the publication year. For somewhat better visual clarity lines have been used instead of step-functions. Countries that are affiliated with ALMA are show in solid lines, countries that are not affiliated with ALMA are listed in {\it italics} and shown with dashed lines. Countries with less than 1\% of the publications in 2024 have been regrouped into `other' (dotted). Chile has a special status as host country and is shown with a dot-dashed line. The countries in the legend are ordered by their publication fraction in 2024.}
  \label{figure:countryfractionevolution}
\end{figure}

Figure~\ref{figure:countryfractionevolution} shows the fraction of first-author countries of ALMA publications as a function of the publication year. Lines have been used instead of step functions for somewhat better visual clarity. ALMA countries are shown with solid lines. Countries that are not affiliated with ALMA are listed in {\it italics} and are displayed in dashed lines. Chile, with its special status, is shown with a dot-dashed line. Countries with fractions of less than 1\% in 2024 are grouped into {\it other} shown as a dotted line. The countries are determined from the location of the first institute the first author is affiliated with at the time of publication.

The decline of the fractions for some of the core ALMA countries like the USA, Japan, or the UK after the initial years is due to the broadening of the author community, including through the accessibility of data for archival research. That said, due to the very strongly rising numbers of publications with time (see Figure~\ref{figure:evolutiongeneral}), in absolute numbers, the number of publications for these core countries has been rising. For instance, the year with the largest number of USA publications was 2023 with 117 publications, Japan in 2022 with 59 publications, and the UK in 2021 with 40 publications.

Beyond the visible increase in publication fraction of the not-listed countries grouped into {\it other}, a notable observation is the first-author publication evolution for non-ALMA countries and countries with modest Gross Domestic Product (GDP) per capita (see below) like Mexico, India and in particular China. China has established itself as the fourth-most publishing country.

A more detailed investigation shows that for 25\% of the Chinese first-author publications, the first author had also a second institute affiliation from an ALMA member state. Also, for all Chinese first-author publications, about 82\% make use of PI data, i.e. data from projects which were proposed by co-authors of the publication (see Section \ref{section:methodology}). In other words, at least some data used was not classified as `archival' usage. This indicates that China is participating strongly in the ALMA science community with significant access to PI data mainly through collaboration with groups from ALMA affiliated countries. We find that roughly three quarters of the datasets that were not based on Open Skies observing time have been published before already by non-Chinese first authors. On average, each of those datasets was published about 2.6 times before a subsequent publication by a Chinese first author. We conclude that due to the high quality of ALMA data, there is additional science contained in the observations which sometimes gets published by Chinese first-authors, who are part of collaborations with the ALMA member state PIs, for mutual benefit and helping maximise the scientific output of the observatory.

ALMA is distributing observing time over the five ALMA regions: CHILE, EA, EU, NA and OTHER (see Section~\ref{section:ALMA}). Investigating the productivity and the community size per region, we find that for all of those regions, if the size of the community as registered in the ALMA Science Portal (see also \ref{subsection:community}) is larger (smaller) than the time-share of that region, then the fraction of publications from that region is also larger (smaller). This indicates that the scientific productivity is more tightly coupled to the number of astronomers that work with ALMA data in a region than to the data-share, i.e. `data do not write papers, people write papers'. Note however that the fraction of publications with multi-region authorship is very large (see Section~\ref{subsection:community}).

\begin{figure}[t]
  \resizebox{\hsize}{!}{\includegraphics{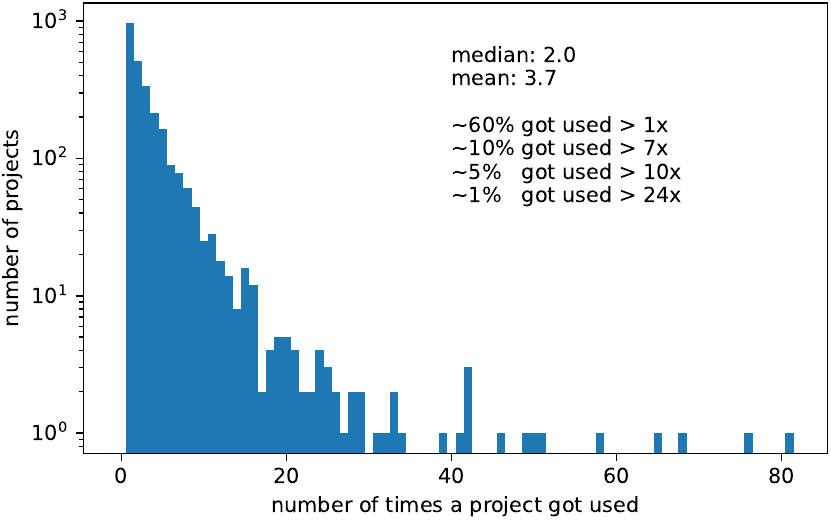}}
  \caption{The number of times published projects have been used in publications between 2012 and 2024. On average each published PI project has been used 3.9 times in different publications. One percent of the PI projects have been used in more than 25 publications.}
  \label{figure:multiple_use_of_alma_projects}
\end{figure}

\subsection{Authors}
\label{subsection:authors}
The fact that ALMA science is becoming more collaborative and more international can also be established by analysing the lists of authors on the refereed publications.

While in 2012 the median and mean numbers of authors on ALMA publications were 6 and 9.9, respectively, these numbers have grown to 10 and 18.9 in 2024. This is still true when excluding publications with more than 200 authors, e.g. publications by the Event Horizon Telescope (EHT) collaboration like \citet{2024A&A...692A.140A_Algaba} with 761 authors, where we find for the mean and median values 10 and 14.6, respectively.

Finally, the fraction of publications written jointly by authors of several ALMA regions has remained remarkably constant between 2012 and 2024: on average 80\% of all publications are multi-ALMA-region publications.

\subsection{Multiple use of data}

ALMA data are very rich and often used in several publications. In Figure~\ref{figure:multiple_use_of_alma_projects} we show the number of projects as a function of how often they have been used in publications for all those PI projects that have been used at least once. Publications that make use of a large fraction of the ALMA archive (ALMACAL series, see also Subsection~\ref{subsection:sciencecategories}), as well as SV, E or CAL projects (see Subsections~\ref{subsection:scienceverification} and \ref{subsection:observingmode}) have been excluded from the analysis. 60\% of the published projects have been used more than once, 10\% more than seven times, 5\% more than 10 times, and 1\% more than 24 times. The median and average values are 2 and 3.7, respectively.

Table~\ref{table:mostoseddatasets} in the Appendix lists the most-used PI datasets in refereed publications, led by 2015.1.00956.S (PI: Adam Leroy) which was used 81 times.

\subsection{Archival research}
\label{subsection:archivalpublications}
\begin{figure}[t!]
  \resizebox{\hsize}{!}{\includegraphics{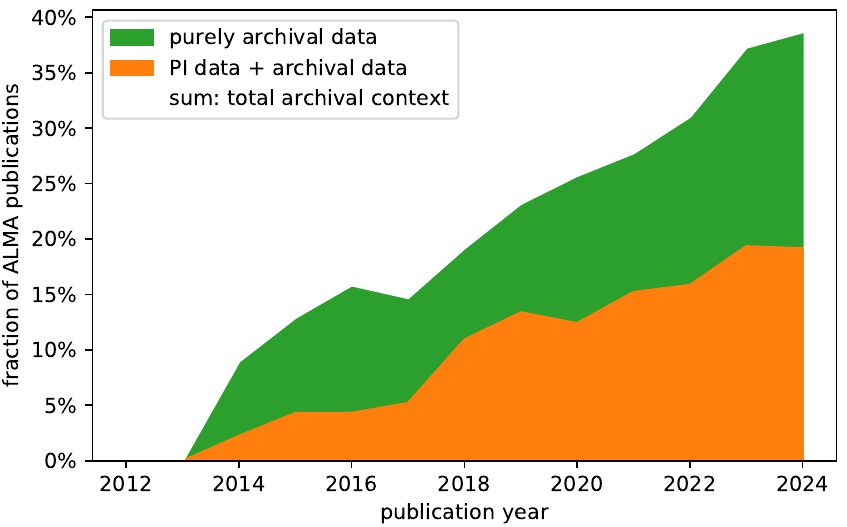}}
  \caption{Fraction of ALMA publications that make either purely use of archival data (green) or of archival data together with PI data obtained by any of the publication authors (orange). SV data have been excluded.}
  \label{figure:archivalevolution}
\end{figure}
All raw data and pipeline products that are delivered to PIs are made publicly available in the ASA after the end of the proprietary period (see Section~\ref{section:ALMA}) and can be used for archival research. The amount of data has been constantly growing and by the end of 2024, the ASA contained about two Petabytes of data from 4,962 projects totalling 81,314 science observations\footnote{An `observation' is defined as a combination of a source and Member OUS} as well as 68,733 calibrator observations. The pipeline products are of `reference image' quality and quality controlled by the observatory. Should problems with the data be detected after delivery, the observatory addresses the problem through the QA3 process\footnote{\url{https://almascience.org/proposing/technical-handbook}} \citep{2020Msngr.181...16_Petry}, after which - in most cases - new products are created and ingested into the ASA.

Figure~\ref{figure:archivalevolution} shows the evolution of ALMA publications making use of data from the ASA. In 2024, a total of 39\% of all ALMA publications utilised at least some archival data. This is a remarkably large value for ground-based facilities, especially only 12 years after the start of observations. To be conservative, we do not count any contribution from SV or Source Catalogue data (as opposed to Figure~\ref{figure:evolutiongeneral}). Also, as mentioned in Section~\ref{section:ALMA}, the measure used is conservative, as any overlap between the author list of the proposal and of the publication would count as PI data usage. An estimate shows, for example, that applying the same method as ESA \citep{2024arXiv240212818D_DeMarchi} where only the inclusion of the PI is used as a criterion, would lead to an archival fraction of more than 50\% for ALMA in 2024. About half of the publications made use of purely archival data (green) and half made use of archival data together with PI data (orange).

We find a substantial fraction of 14\% of published projects that have never been published by the PI group but have been published by archival researchers, even if we exclude publications like the ALMACAL series. Remarkably, a total of 13\% of all publications in the high-impact journals (see Subsection~\ref{subsection:journals}) {\it Nature} and {\it Science} make use of ALMA archival data, and five publications (6.4\%) in those two journals exclusively use archival data. We confirm that the availability of archival data for research massively increases the scientific productivity of the observatory.

\begin{figure}[!t]
  \resizebox{\hsize}{!}{\includegraphics{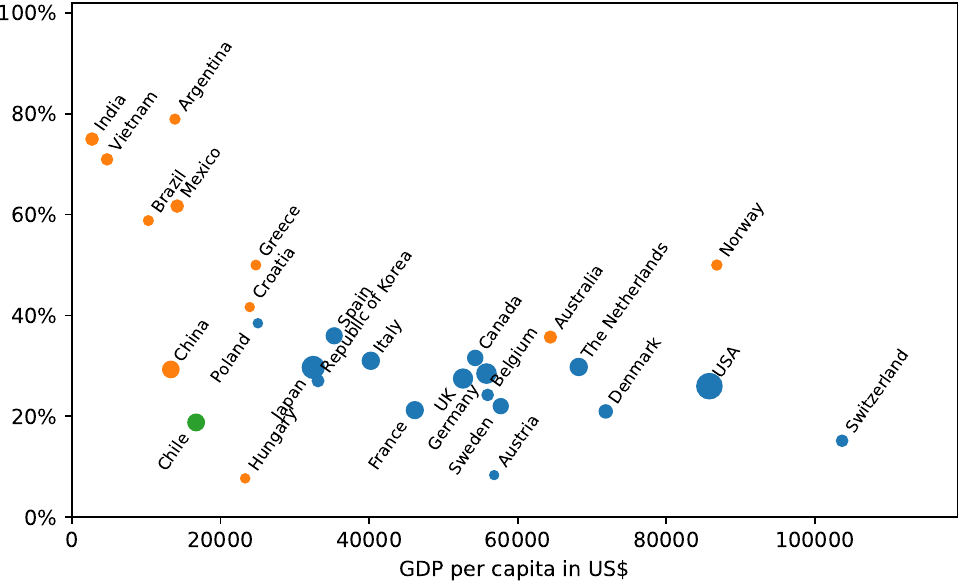}}
  \caption{Average archival fraction of datasets used in each publication as a function of the GDP per capita of the respective first-author country. ALMA countries are shown in blue, the host country Chile is shown in green and other countries are shown in orange. The surface area of the markers is proportional to the number of publications of that country.}
  \label{figure:archivalfractionvsadjustednetnationalincomepercapita}
\end{figure}

\citet{2019BAAS...51g.105P_Peek} show that archival research can enable science by astronomers from countries with more modest GDP per capita even if those astronomers have in principle access to telescope time through Open Skies (see Section \ref{section:ALMA}). We show the results of an equivalent analysis for ALMA in Figure~\ref{figure:archivalfractionvsadjustednetnationalincomepercapita}. As a function of the GDP per capita\footnote{\url{https://data.worldbank.org/indicator/NY.GDP.PCAP.CD}, values mostly from 2024 expressed in current US\$.} in current US\$, we show the average fraction of the publications that make use of any archival data. Only countries that have at least ten publications attributed to them are plotted. ALMA countries are shown in blue, non-member countries in orange, and the host country Chile in green. The surface area of the markers is proportional to the number of publications of that country.

We find a very clear trend - in line with \citet{2019BAAS...51g.105P_Peek} - of astronomers in countries like India, Vietnam, Brazil, Argentina, and Mexico with less than 15 kUS\$ of GDP per capita engaging very strongly in archival research, while on average publications with first authors from ALMA member state countries predominantly are based on PI data. A notable exception is Chile, which has a low archival fraction due to the substantial share of 10\% of the ALMA observing time. Hungary, Norway, Australia, and China (see Subsection~\ref{subsection:countries}) have lower average shares due to strong collaborations of astronomers in those countries with astronomers in ALMA countries. Archival research not only enables more science in general but also in particular broadens the ALMA user base beyond ALMA countries in addition to the access to Open Skies data.

\subsection{Publication delay and rate}
\label{subsection:publicationdelay}
\begin{figure}[t!]
  \resizebox{\hsize}{!}{\includegraphics{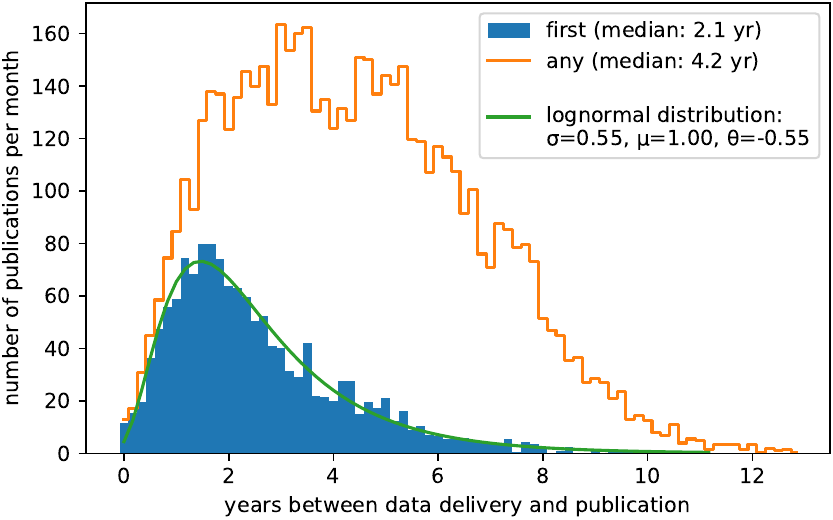}}
  \caption{Distribution of the time between the first (blue) or any (orange) publication as a function of the time between the delivery of the data to the PI group and the publication. The data have been smoothed into 2-month bins. The green line shows a fit of log-normal distribution to the first-publication distribution.}
  \label{figure:publicationdelay}
\end{figure}
Astronomy is a fast-paced and ever-growing field of science. By now ADS lists nearly 100 new refereed publications per day for `astronomy' collection alone. The success of an observatory - as we have argued before \citep{2022Msngr.187...25S_Stoehr} - is closely related to the ease and speed with which astronomers can reach results they need and how much they can actually concentrate on doing science as opposed, for example, on combining data from the different ALMA arrays into a single FITS product. We have coined the concept {\it fastronomy} for this need \citep{2022Msngr.187...25S_Stoehr}. In Figure~\ref{figure:publicationdelay} we can now measure directly the time it takes from the delivery of data to the PI group and the first (blue) or any additional publication (orange) by that group. I.e., we explicitly do not include any archival research in this analysis. For all calculations in this work, we use the median value of the Member OUS delivery dates of a program as the delivery date of that program.

It takes the PI and collaborators a median of 2.1 years (mean 2.55 years) until the first publication appears, which is consistent with  the findings for XMM \citep{2025AN....34670014_Ness}. It takes a median of 4.2 years (mean 4.41 years) for any publication they write. This time is significantly longer than the proprietary period of 12 months (or six months for DDT projects). It is also a large fraction of (or may even exceed) the time it takes to obtain a PhD or the length of a post-doc contract. The time-to-publication has also increased significantly from the 1.4 years of ALMA's Cycle 0 projects \citep{2015Msngr.162...30S_Stoehr}. In extreme cases, it takes a group over ten years to publish their data for the first time.

The green line shows a fitted log-normal distribution with $\sigma=0.55$ (standard deviation), $\mu=1.0$ (mean value of the distribution) and $theta=-0.55$ indicating that the distribution is Gaussian in log-time between delivery and publication. It is remarkable that in both cases, the first publication astronomers write {\it after} data delivery, as well as the proposal submission times {\it before} an ALMA deadline \citep{2017msngr.169...53_Stoehr}, are Gaussian distributions - but with a logarithmic time axis. Log-normal distributions arise in random processes where the random variable is a product of random variables \citep[e.g.][]{Limpert2001Lognormal}.

While it is natural that, as projects become larger and more complex, and as publications increasingly rely on data from several wavelength regimes and facilities, the publication delay increases, it nevertheless seems important to evaluate the astronomer experience carefully, identify the obstacles, and address them \citep{2022Msngr.186...20H_Hatziminaoglou}. The time to publication needs to be short enough to be in line with the scientific progress in the field, or data will remain unused and unpublished (see also Subsection \ref{subsection:publicationfraction}). This is not only important to maximise the scientific output of the facility by allowing scientists to spend their time more effectively. It is also important to keep ALMA competitive for graduate students who may see faster and larger potential, orienting themselves towards a facility like Gaia - with its very high-level and comparatively low data size catalogue - when choosing their career path.

\begin{figure}[t]
  \resizebox{\hsize}{!}{\includegraphics{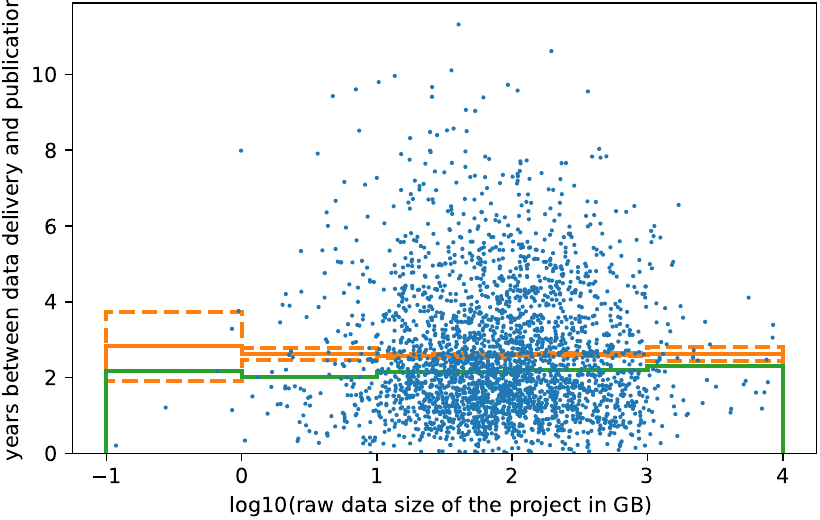}}
  \caption{Distribution of the delay between the data availability to the PI (or archival researchers) and the first resulting publication as a function of the size of the raw data of the observing program. The mean and standard uncertainty are shown in orange colour with solid and dashed line styles, respectively. We find there no indication for a correlation e.g. that smaller projects would get published faster. The median is shown in green and indicates a slight trend.}
  \label{figure:publicationdelayvsdatasize}
\end{figure}

One hypothesis frequently put forward is that the publication delay has increased due to the larger size of data delivered to PIs, which is a consequence of the increased number of antennas available for observing since Cycle 0 and the encouragement from ALMA for PIs to propose longer projects. In Figure~\ref{figure:publicationdelayvsdatasize}, we show the time it takes the PI group to publish the first publication of a particular project as a function of the project's total raw-data size (a more reliable measure of the project data volume than the size of the products, which ALMA does not necessarily produce in its entirety). When looking at the average values shown by a solid orange line with the uncertainty interval plotted in dashes, we see no correlation and thus no confirmation of the hypothesis. Over five orders of magnitude in project size, on average, data from small projects is published for the first time as rapidly as medium-sized or large projects. It takes an average of 2.65 years\footnote{The difference to the 2.55 years in Figure~\ref{figure:publicationdelay} is due to the averages being once computed per publication and once per project.} for any given project to appear for the first time in a publication by the PI group. The green line shows the median values for the same bins where a small trend is observable. E.g. projects of raw data sizes one to 10 GB are typically (but not on average) published roughly 3.6 months faster than projects of sizes between one and 10 TB.

\begin{figure}[t]
  \resizebox{\hsize}{!}{\includegraphics{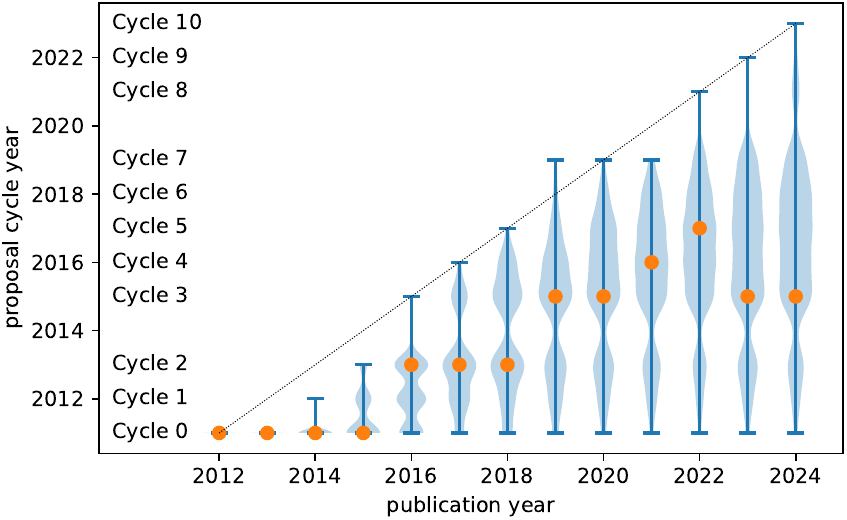}}
  \caption{Distribution of the proposal cycle year of data that were used in articles as a function of the year the article was published. The doted line indicates the year of the first possible publication for the data of a given proposal cycle. Orange dots show data from which cycle was the most popular in the publications of a given year.}
  \label{figure:whichdatagetspublishedmosteachyear}
\end{figure}
It is already clear from Figure~\ref{figure:publicationdelay} that ALMA data are relevant for a very long time. Even in 2024, authors have made use of their PI data from 14 projects of Cycle 0. The longest delay between data delivery and the very first publication by the PI group using those data recorded so far is 10.5 years.

We now examine more closely the distribution (blue) of the data of the various ALMA proposal cycles in the publications appearing in a particular year in Figure~\ref{figure:whichdatagetspublishedmosteachyear}, this time including data used in archival research. We use years instead of cycles for the vertical axis to account for the fact that not all ALMA cycles were exactly one year long. The year of the start of observing was used for each cycle. The grey line is a guide for the eye plotted to indicate publications that would happen in the year following the start of observations, i.e. a rough indication of the proprietary time. Orange dots indicate the cycle from which the largest number of projects were used in publications of the given year.

Publications between 2012 and 2015 predominantly made use of Cycle 0 data. With the opening of a new window to the Universe (see also Subsection~\ref{subsection:journals}), substantial scientific progress was made from those data. At the same time, ALMA's capabilities  increased, and by Cycle 2, the number of 12-m antennas typically used had nearly doubled from 19 to 36. Cycle 0 and Cycle 1 observations dropped in popularity, with Cycle 2 dominating publications up to 2018. For publications from 2018, the most relevant data are those from Cycle 3 onward. Two effects are highly remarkable: first, Cycle 3 data remain the most published data even in publications of 2023 and 2024. And second, data from the very recent cycles are far less used in recent publications than data from Cycles 3 to 7.

While still used in some publications even in 2024, data from Cycles 0 to 2 have essentially been superseded by observations of later cycles with better ALMA capabilities - often re-observing the same sources. Data from Cycle 3 onward however, remain highly relevant to extract new science, and no decline is observable. The low usage of data from the very recent cycles will need to be monitored carefully going forward.

\subsection{Publication fraction}
\label{subsection:publicationfraction}
\begin{figure}[t]
  \resizebox{\hsize}{!}{\includegraphics{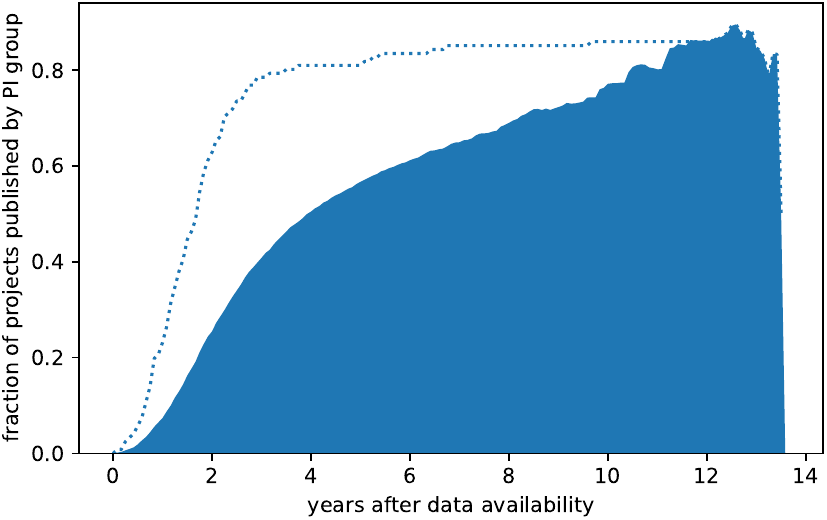}}
  \caption{Evolution of the publication fraction of PI-data by the PI-groups as a function of the years the data have been made available to them. The dotted line shows the same evolution but for Cycle 0 data.}
  \label{figure:publicationfraction}
\end{figure}
The ALMA proposal process is highly competitive, with typically about four to seven times more observing time requested than can be accepted in any given cycle\footnote{\url{https://almascience.org/news/documents-and-tools/cycle10/cycle10-proposal-process}}. Substantial effort is deployed to select the science with the largest scientific merit \citep{2023prur.confE..17C_Carpenter}, and observations are then carried out to the specifications requested by the successful PIs.

On average, considering data observed between 2011 up to and including 2021 as well as publications up to and including 2024, we find a publication fraction for ALMA of $\sim$60\% for data published by the PI group, and $\sim$70\% for data published by PIs as well as by archival researchers.

Overall, up to the end of 2024, we find that in terms of observing time, out of the total of 37,375 hours of 12-m-equivalent time spent on all projects together, 19,362 hours, i.e. 52\%, have actually been used in publications. This is consistent with the fact that out of the 4,929 PI projects observed so far, 2,639 have been published, i.e. 54\% (See Table~\ref{table:sciencecategories}).

Figure~\ref{figure:publicationfraction} shows the fraction of the observed ALMA projects that were published by the PI group (filled), i.e. how effectively the data were converted into science. It takes on average about four years until half of the data are published by the PI group. The dotted line shows the evolution for Cycle 0 data alone. The projects of the first cycles were published to a higher fraction far more rapidly, with half of the projects published in less than two years.

\begin{figure}[t!]
  \resizebox{\hsize}{!}{\includegraphics{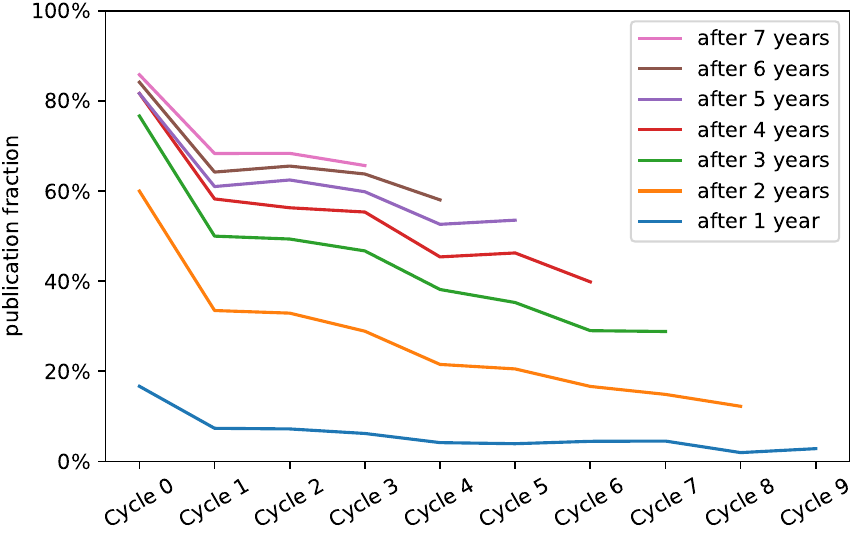}}
  \caption{Evolution of the publication fraction of PI-data by the PI-groups - in years after the data have been made available to them - as a function of ALMA's observing cycles. For better visual clarity continuous lines instead of step-functions have been used. }
  \label{figure:evolutionpublicationfractioncycle}
\end{figure}
We observe two effects acting simultaneously which cannot be disentangled: 1) The fraction of data that are published by the PI group is decreasing as the observatory matures. 2) The delay between data delivery and publication by the PI group increases over time. In other words, if projects are not yet published by the PI groups, we cannot know if the reason is the increasing delay in publication (see Figure~\ref{figure:publicationdelay}) or whether the data will never be published by the PI group.

We examine the publication fraction evolution more closely in Figure~\ref{figure:evolutionpublicationfractioncycle} where we show the PI-group publication fraction as a function of the ALMA cycle in which the data were observed. The lines - which we use instead of step functions for better visual clarity - show the evolution after a given number of years. Over 15\% of the Cycle 0 data (project code 2011.0) were published by the PI group within one year, and after three years over 75\% of the projects were used in publications. This evolution has dropped significantly since then, and the conclusion from
\citet{2015Msngr.162...30S_Stoehr}, that 80\% of the data are published, is only true for Cycle 0 data. After three years, only about 30\% of the Cycle 7 data (project code 2019.1) have been used in publications from the PI group. And less than 3\% of all projects of Cycle 9 (project code 2022.1) were published within the 12 month proprietary period.

There is no obvious plateau reached for the evolution of the publication fractions; although one might argue that the one- and five year lines may start to show one. A plateau would indicate that a stable publication fraction has been reached. We rather have to conclude that as time marches on, both, the fraction of data that gets published by the PI group continues to decrease and the time it takes to publish the results from ALMA projects continues to increase. This is a known pattern for maturing observatories \citep[e.g.][]{harwit1981cosmic}. But we conclude at the same time from the evolution in the vertical direction in Figure~\ref{figure:evolutionpublicationfractioncycle} that the publication fraction by the PI groups will continue to increase, as it even increases from the 6th to the 7th year for data from all cycles.

\begin{figure}[t!]
  \resizebox{\hsize}{!}{\includegraphics{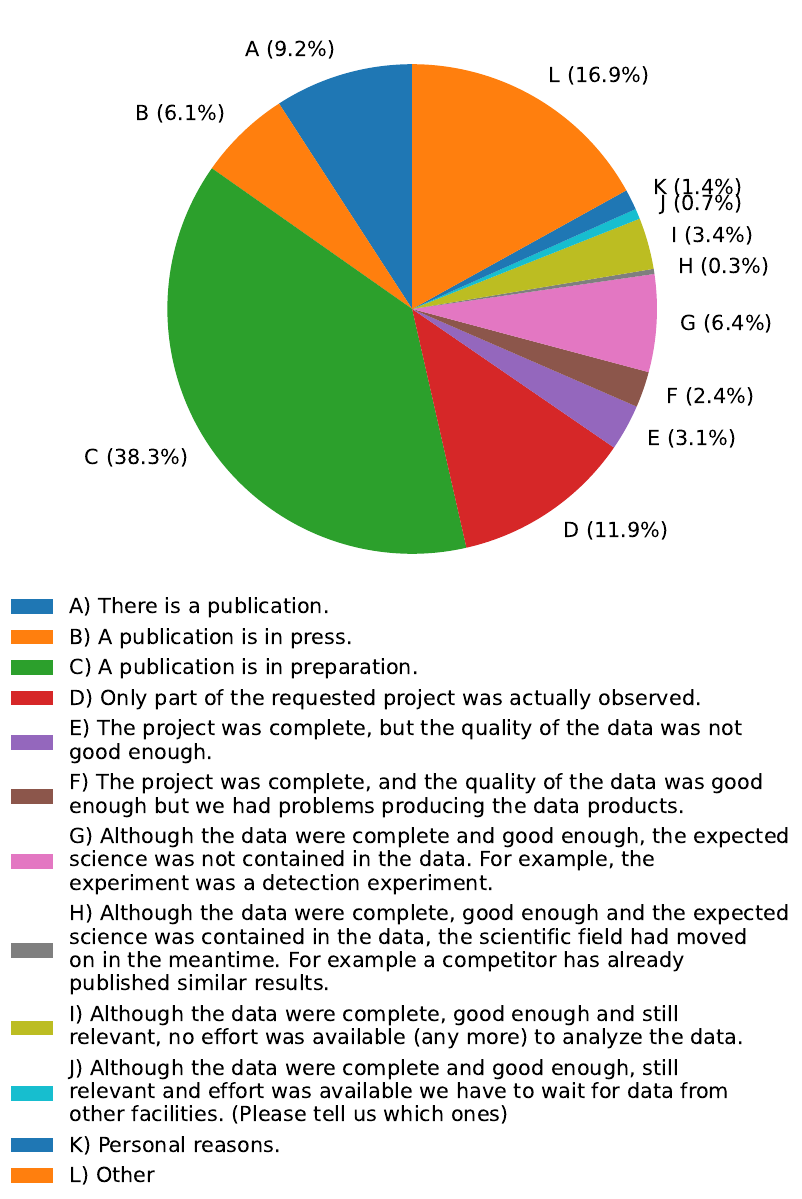}}
  \caption{Answers to the anonymous questionnaire sent to PIs two - since 2023 three - years after the data-delivery if no corresponding publication could be identified. In 38\% of cases, PIs are still working on a publication. Remarkably, only 6\% of the respondents indicated that the expected science was not contained in the data. Only 0.3\% of the PIs indicated that the data were not relevant any more.}
  \label{figure:unpublishedsurvey}
\end{figure}

The fraction of projects that can be published by the PI group as intended is directly impacting the success of the observatory. It is expected that not all observations, e.g. detection experiments, will lead to publishable results. Still, the publication fraction is an important measurement for the effectiveness of the end-to-end science operations, and the observed evolution should be followed up and the observatory should strive to continuously improve operations to further facilitate the extraction of science from ALMA data.

As mentioned in Section~\ref{section:methodology}, to investigate the status of the projects and to gain more insight into the reasons why PI-groups were not able to use their data for publication, ALMA continuously runs an anonymous questionnaire in case no publication could be identified after two to three years \citep{2016arXiv161109625S_Stoehr}. Figure~\ref{figure:unpublishedsurvey} shows the statistics of this survey, which was filled out by 22\% of the 1366 PIs notified. Category C,  `A publication is in preparation', clearly dominates with 38\% of the answers. An unexpectedly low value of six percent of the respondents indicates that the expected science was not contained in the data, as such, but also compared to e.g. 19\% for the VLT \citep{2017msngr.170...51P_Patat}. This shows that improvements to science operations have the potential to significantly increase the publication fraction and thus the scientific impact of ALMA: A total of 26.8\% of the cases are related to the quality or completeness of the data, no time being available any more or `other' reasons. A very small fraction of 0.3\% indicated that the science is not relevant any more, e.g. a competitor has published similar results.

\subsection{Citations}
\begin{figure}[t]
  \resizebox{\hsize}{!}{\includegraphics{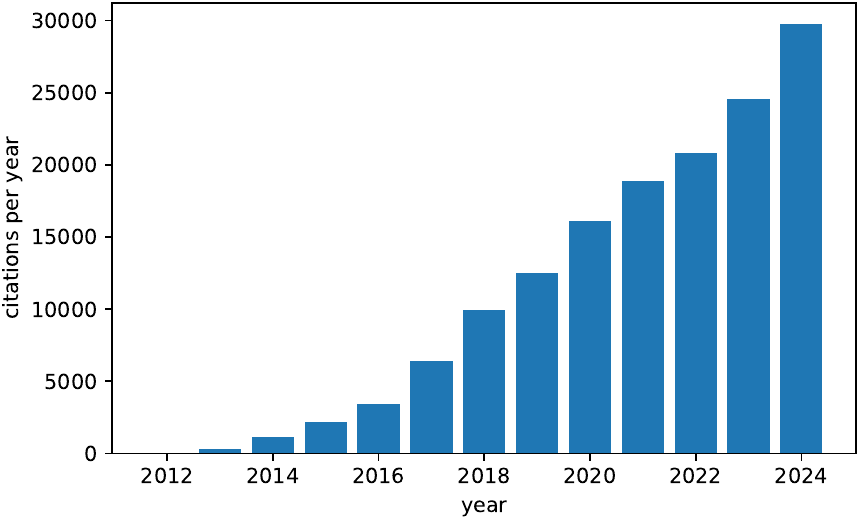}}
  \caption{Number of citations received by all ALMA publications per year.}
  \label{figure:citationsperyear}
\end{figure}

The second measure that is most often used to evaluate the success of an astronomical facility is the number of citations that the publications receive - often referred to as the scientific impact. We use here the citations as tracked by ADS. These citations include only references from refereed publications.

The total number of citations is shown in Table~\ref{table:journals} alongside the number of publications and the citations per hour of observing time for each scientific category. Table~\ref{table:receiverbands} and Figure~\ref{figure:publications_citations_per_band} provide information about the citations received in each receiver band. Those numbers have been discussed in their respective sections.

\begin{figure}[t]
  \resizebox{\hsize}{!}{\includegraphics{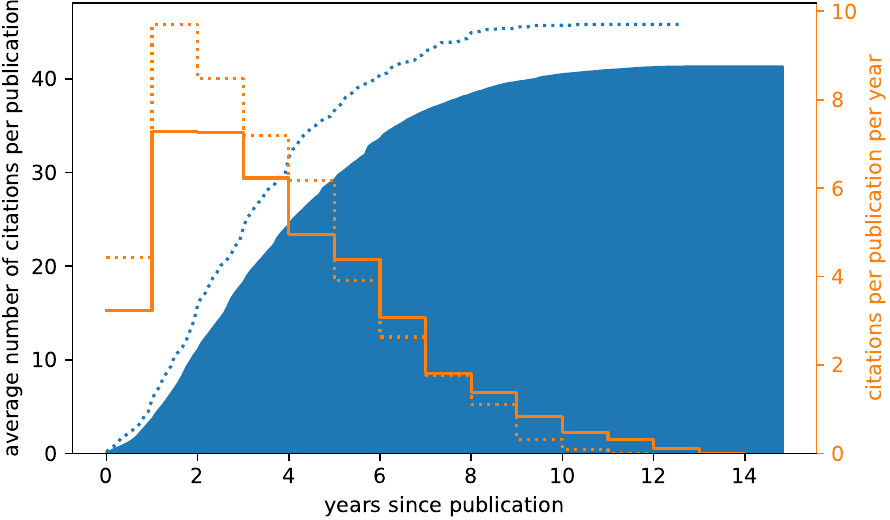}}
  \caption{Average cumulative number of citations per publication after a given number of years (blue) and the yearly average number of citations per publication (orange). Dotted lines show the values for Director's Discretionary Time data (DDT).}
  \label{figure:citationsperpublication}
\end{figure}

Table~\ref{table:largestcitationrate} in the Appendix lists the 50 ALMA publications with the largest citation rate. Following \citet{2014AN....335..210N_Ness}, we do not show publications from 2024 to eliminate short-term effects. This list, dominated by publications from the EHT Collaboration, would remain similar had the table been sorted by the total number of citations instead of by citation rate. ALMA's contribution to the EHT VLBI network is pivotal due to its provision of the largest and most sensitive collecting area, its favourable geographical location, and its shorter baselines, which together enhance the completeness and quality of the spatial sampling of the array, thereby significantly improving the overall imaging capability and performance.

Figure~\ref{figure:citationsperyear} shows the number of citations that all ALMA publications are receiving together per year. The number of citations is rising very rapidly, more rapidly than the number of publications (see also Section~\ref{section:comparisonwithotherfacilities}). Figure~\ref{figure:citationsperpublication} shows the cumulative average number of citations of ALMA publications (blue) as a function of age of the publication for all PI data (filled solid) as well as the yearly average value (orange solid line). 

\begin{figure}[b]
  \resizebox{\hsize}{!}{\includegraphics{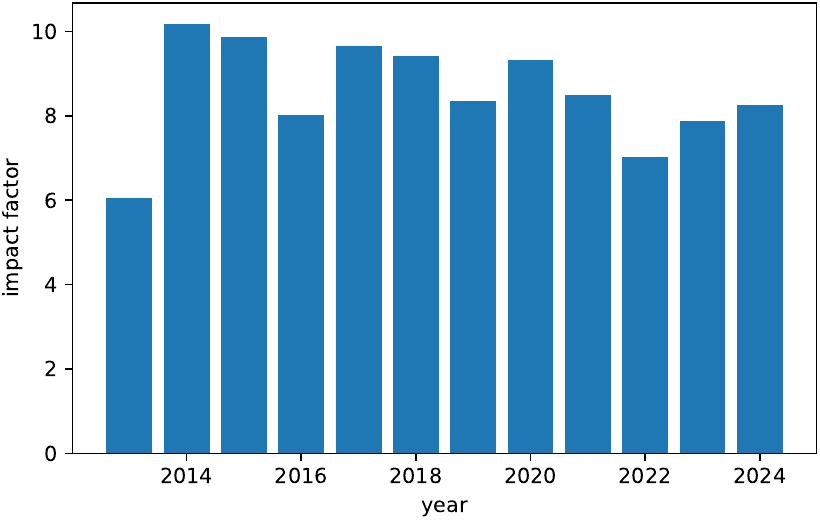}}
  \caption{The impact factor - defined as the number of citations in a given year resulting from the publications from the previous two years - as a function of the year. The impact factor is relatively constant over time. On average each publication between one and two years old has received about eight citations in the previous year.}
  \label{figure:impactfactor}
\end{figure}

Dotted lines show the average cumulative evolution (blue) as well as the yearly average evolution (orange) for the subset of publications that make use of at least some DDT data. Publications making use of DDT data receive more citations initially as well as more citations overall compared to all ALMA publications.

\begin{figure*}[t]
  \resizebox{\hsize}{!}{
  \includegraphics{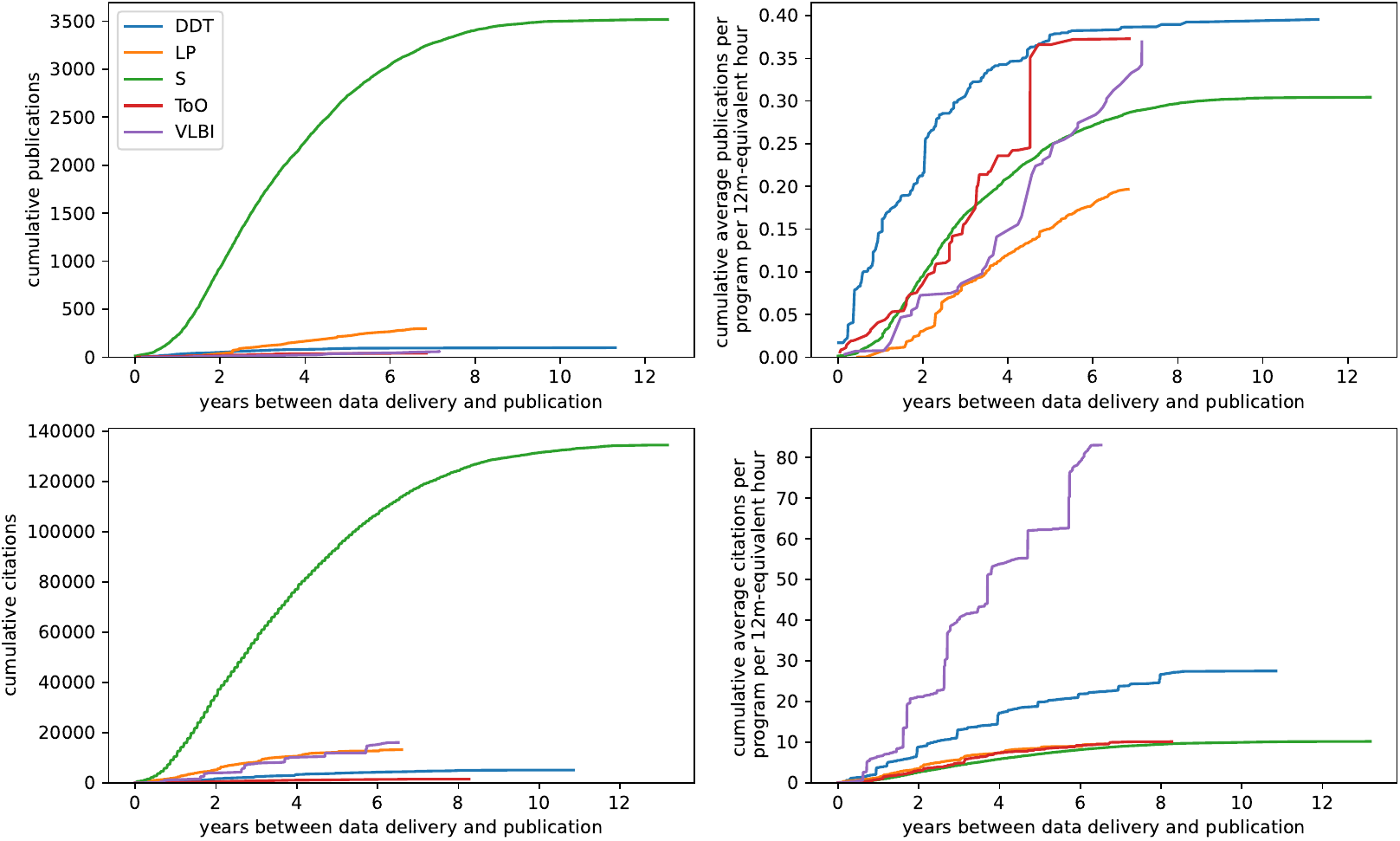}}
  \caption{Evolution of the cumulative average number of publications (top) and citations (bottom) in absolute values (left) and normalised per program and per 12-m-equivalent observing time (right). The calculations are made per published ALMA program and given as a function of the time between the delivery of the program and its use in one (or several) publications.}
  \label{figure:observingmode}
\end{figure*}

\subsection{Impact factor}
\label{subsection:impactfactor}
A third measure of scientific productivity, beyond the number of publications and the number of citations, is the impact factor, defined as the total number of citations received in a given year {\it stemming from publications of the previous two years}\footnote{\url{https://en.wikipedia.org/wiki/Impact_factor}}.

We show the impact factor as a function of the publication year for ALMA publications in Figure~\ref{figure:impactfactor}. Despite the evolution of ALMA from an entirely new facility to a mature observatory, and despite the rapid evolution of both the number of publications as well as the number of citations over time, the impact factor remains remarkably stable. However,  a slight decrease of about 19\% from, e.g. 10.2 in 2014 to 8.25 in 2024 is noticeable. Averaged over the last five years, ALMA's impact factor is 8.21. We compare the impact factor with that of other facilities in Subsection \ref{section:comparisonwithotherfacilities}.

\subsection{Science Verification}
\label{subsection:scienceverification}
Each time a new capability is offered, ALMA releases corresponding SV data to the public without a proprietary period. Especially in the very first years of ALMA operations, in addition to familiarising the community with ALMA data, these data were a major contributor to the scientific productivity of the observatory. In 2012, 63\% of the publications made use of SV data. The fraction dropped to 30\% in 2013 and 11\% in 2014. With the publication of \citet{2015ApJ...808L...3A_ALMA} and the corresponding SV data release of 2011.0.00015.SV, the fraction rose temporarily to 23\% again before rapidly dropping to levels of about 1\% in recent years. This evolution is tightly coupled to the releases of SV data by the observatory, which happened very frequently in the first years when many new capabilities were brought online\footnote{\url{https://almascience.org/alma-data/science-verification}}. Up to and including 2024, 193 publications made use of SV data.

The median time between the availability of SV data and the first related publication is less than 5 months, i.e. about 1/5th of the time for typical first publications of PI data. The median time until {\it any} related publication is 1.9 years and thus less than half the timespan of typical ALMA projects (see Subsection~\ref{subsection:publicationdelay}). The most popular SV data delivery is 2011.0.00009.SV from 2012 containing data of Orion KL which was used in 34 publications (see Table \ref{table:mostoseddatasets} in the Appendix).

\subsection{Observing Modes}
\label{subsection:observingmode}
ALMA offers several observing modes. The Standard mode is the default and dominates the number of approved ALMA programs. Since Cycle 4 in 2016, astronomers can submit Large Program (LP) proposals that ask for more than 50 hours of observing time on the 12-m array. In order to be accepted, Large Programs need to receive a Grade~A in the proposal review process, which also means that if they are not fully observed in the cycle they were accepted in, they get carried over into the next cycle for additional observations. Astronomers also have the possibility to propose for Target of Opportunity (ToO) projects where the science is laid out in the proposal, but the target and/or time of the observation are not known in advance - e.g. a comet, a supernova, or a gamma-ray burst. The ALMA Director has the discretion to allocate up to 5\% of the available time during a cycle as DDT. DDT proposals can be submitted any time during a cycle and have so far typically been used to follow up on unexpected high-interest discoveries from other facilities or from ALMA itself. Finally, ALMA offers the VLBI mode in which observations are coordinated and combined with observations from other facilities worldwide through the EHT and the Global Millimeter VLBI Array (GMVA) collaborations. Details of these observing modes are provided in \citet{alma_proposers_guide2025} and \citet{alma_technical_handbook2025}. 

Assigning to each publication the project type that was predominantly used, we find that for the publications of 2024 the project type distribution - as defined by the observatory and added to the end of each ALMA project code, or for DDT indicated by an A in the cycle number - was (S)tandard 84\%, (L)arge Program 8.3\%, DDT 2.5\%, (V)LBI 2.3\%,  (T)arget of Opportunity 1.1\%,(S)cience(V)erification 0.72\%, (E)ngineering 0.72\%, and (CAL)ibrator at 0.54\%.

Figure~\ref{figure:observingmode} shows a comparison of the ALMA observing modes for the cumulative number of publications (top) and the cumulative number of citations (bottom) as a function of the time a project has been observed and its use in a publication, per project type. The citations a publication has received are distributed over the projects used in that publication by the fractional 12-m-equivalent observing time of the projects. The left column shows the distribution in absolute values and the right column normalised per program and per 12-m-equivalent observing hours. SV, E, and CAL programs have been excluded from the calculations.

Standard programs dominate the absolute number of publications and citations (left). The relative productivity of the different observing modes can be inferred from the right hand side of Figure~\ref{figure:observingmode}. Normalised by program and by the 12-m-equivalent observing time ALMA has invested, the observing modes produce a roughly similar amount of publications with DDT programs resulting in a slightly larger and LP a slightly lower number than the average Standard project. We attribute the LP curve to the fact that LPs typically create high-level products of large value for the community \citep[e.g.][]{2018ApJ...869L..41_Adrews, 2021ApJS..257...43_Leroy} which then get used directly instead of the original ALMA data. According to our methodology (see Section~\ref{section:methodology}), such derived works however are not counted as ALMA publications. As those derived works do however cite the original LP publications, the evolutions of the citations normalized per program and per invested 12-m-equivalent observing hour shown in the lower right panel are then remarkably similar for LP, ToO and Standard observations. It might be that once a longer time-line is available, the higher legacy value of LPs may lead to an increased number of citations per invested hour. DDT programs - related to their typically highly relevant new results - get cited significantly more often. 

We note that the largest amount of citations normalised by observing time is by far received by VLBI programs normalised by the observing time\footnote{By construction, VLBI observations are made from data of several facilities observing the same source quasi-simultaneously. For the normalisation here, only the ALMA observing time is accounted for.}.

\begin{figure}[t]
  \resizebox{\hsize}{!}{\includegraphics{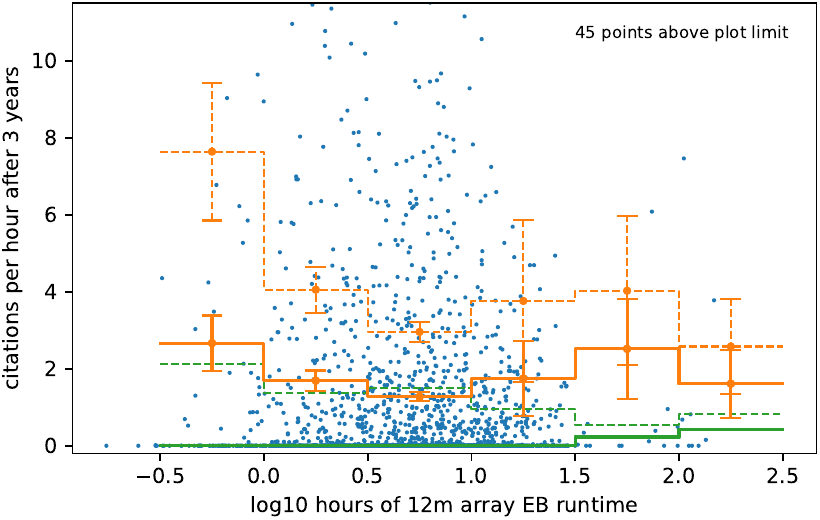}} \caption{Number of citations accumulated in the first three years after the publications, distributed over the ALMA programs and bands actually used in the publications weighted by their observing times (blue dots where for visual clarity the plot limit has been set so that 45 dots are not shown), averaged and shown as as a function of the invested ALMA observing hours of the projects. Only the values for the 12-m-array are shown and we only take into account projects after 2016 when LPs were first introduced to not introduce a bias. The average number of citations per hour of invested time of all observed projects is constant (orange solid) over three decades of observing time. We include in this measurement also projects that have not been published and thus have received no citations. Looking only at the projects that have actually been published (orange dotted), there is a very marginal signal for projects with less than one hour of observing time to generate more citations per invested hour. The difference is due to the bias in the the publication fraction, where the larger the project is, the more likely it is to receive at least one publication. The same effect can be observed when looking at the median values (green) of the two measurements. When all observed projects are taken into account there is a very slight indication for larger numbers of citations for the typical program exceeding 30 hours of observing time, but this trend is reversed when only taking into account the projects that have been actually published for the 'per hour' normalisation, where a clear increase of the number of citations for shorter projects can be observed.}
  \label{figure:citationsperhourvsarraytimeperprogram}
\end{figure}

As by construction Figure~\ref{figure:observingmode} only analyses programs that have been published, we here give the average publication fraction per observing mode for data observed until the end of 2021 published by the PI group or through archival research. We find publication fractions of 85\%, 69\%, 68\%, 59\%, and 54\% for LP, Standard, DDT, ToO, and VLBI, respectively.

We have concluded above that LPs on average produce the same number of citations per program, normalised per 12-m-array-equivalent observing time as e.g. Standard programs. In Figure~\ref{figure:citationsperhourvsarraytimeperprogram} we now have a more detailed look into the productivity of ALMA programs as a function of their logarithmic 12-m-array-equivalent observing time. Only the values for the 12-m-array are shown for clarity but the 7-m and TP figures look similar. The citations received are measured three years after a publication appeared and are again distributed over the programs according to their observing time fraction. We restrict the analysis to projects observed after 2016 where LPs were offered for the first time to avoid a bias. Over three decades in total EB runtime of a project we find no signal in the received citations when normalised by observing time (orange solid). For this measurement also projects that have not been published at all and have thus received no citations are taken into account (blue dots for the individual measurements). When we look only at the projects that actually have been published (orange dashed) we do find a very marginal signal for very short projects of less than one hour of observing time receiving more citations per hour than larger projects. The reson for this difference is the bias that the fraction of projects that receive a publication is larger for programs with larger amounts of observing time. Whereas a very short project may be a high-risk but high-reward experiment with the possibility that the expected science is not contained in the data, it is virtually guaranteed that any Large Program with more than 100 hours of observing time will result at the very least in a few publications. The same finding is confirmed when analysing the median (green) instead of the average of the distributions. When taking into account published and unpublished projects (green solid) there is a slight trend for long projects typically receiving somewhat more citations than short projects, whereas the trend is more pronounced and reversed when concentrating only on projects that have actually been published (green dashed).

\begin{figure}[t]
    \centering
     \resizebox{\hsize}{!}{\includegraphics{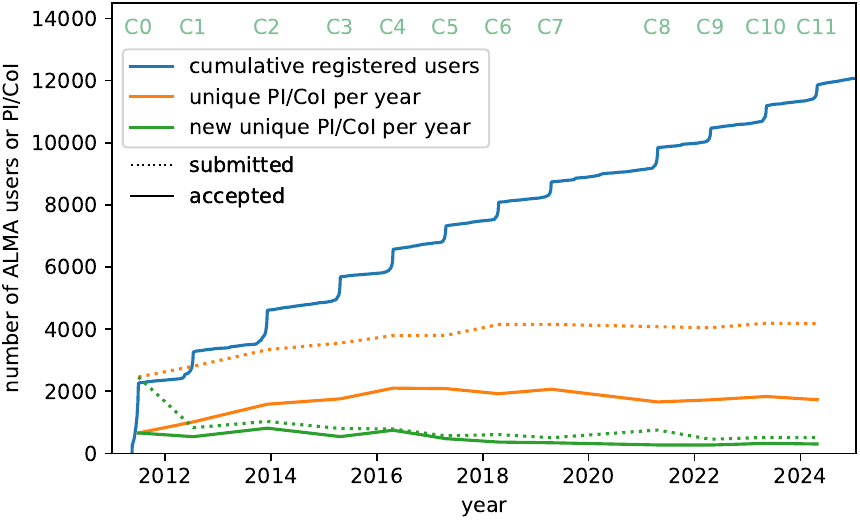}}
    \caption{Evolution of the number of registered users in the ALMA Science Portal (blue). Note that all authors of ALMA proposals need to be registered. At the end of 2024, 12,158 users were registered. A linear regression fit excluding the initial registration for Cycle 0 yields a slope of 740 new users registering each year. Overplotted are the number of unique users on ALMA proposals each year (orange) and the number of new unique authors (green) - i.e. authors that never had been on a proposal before. Dotted lines indicate the submitted proposals and solid lines the accepted proposals of Grades A and B.}
    \label{figure:useraccountstatistics}
\end{figure}

We can derive two conclusions from this analysis. First, the data do not support arguments for increasing or decreasing the amount of observing time that is attributed to very long programs like Large Programs. Per invested hour of observing time the amount of citations received is essentially independent of the length of the program. And second, that if it was possible to improve science operations so that an increased fraction of shorter projects lead to a publication, then allocating more of the total observing time to smaller projects would increase the total amount of citations ALMA is expected to receive.

\subsection{Community}
\label{subsection:community}

ALMA was designed to be a millimetre/submillimetre facility used not only by radio interferometry experts but by all astronomers instead \citep[e.g.][]{2008SPIE.7012E..0NH_Hills}. A way to track the evolution of the ALMA user community is by looking at the number of user accounts in the ALMA Science Portal. This number is a robust indicator, as registration with the ALMA Science Portal is mandatory for each PI and CoI prior to the submission of an ALMA proposal.

Figure~\ref{figure:useraccountstatistics} shows the number of active user accounts as a function of the year. By the deadline of the first call for proposals in 2011, 2,252 users had registered. We speculate that this number encompasses essentially the entire millimetre/submillimetre community at that time.

Since then, the number of registered users has been growing nearly linearly over time at an average rate of about 740 new astronomers per year. At the end of 2024, a total of 12,073 users had registered, which is a factor of 5.4 larger than at the Cycle 0 deadline. We see this as direct evidence for the growth of the ALMA community far beyond the original millimetre/submillimetre community and thus the fulfilment of ALMA's promise. 

In 2024, about one third of all registered users were active and authors on ALMA proposals. Averaged over the last five years, about 45\% of all unique PI/CoI per year are on at least one proposal that got accepted. For first-time PI/CoI, the ratio is not smaller but larger: 56\% are on a proposal that has been accepted.

\begin{figure}[t]
    \centering
     \resizebox{\hsize}{!}{\includegraphics{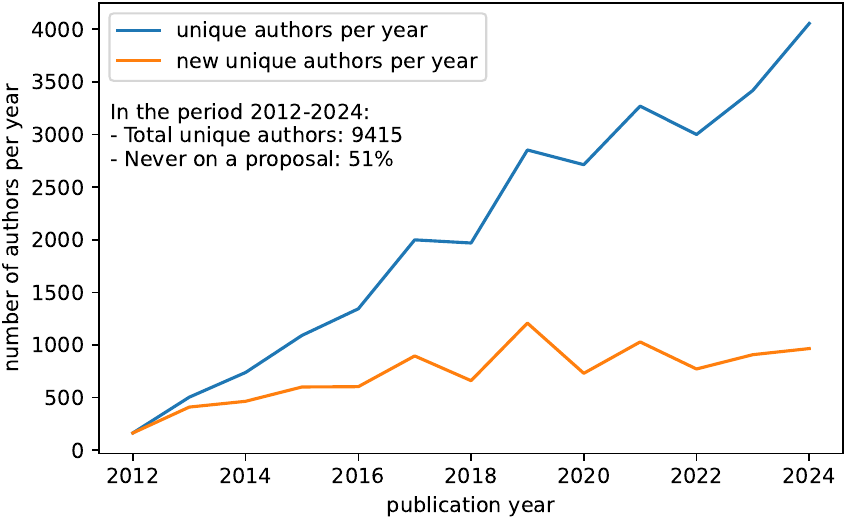}}
    \caption{Analysis of the number of authors per year on ALMA publications as a function of the publication year. We use an algorithm to determine the most plausible combination of the last name and the first letter of the first name to estimate a lower bound of unique authors. There are about 880 new unique authors newly appearing on ALMA publications each year in the last five years.}
    \label{figure:publicationauthoranalysis}
\end{figure}

We have also carried out an analysis of the number of authors on ALMA publications as a function of the publication year in Figure~\ref{figure:publicationauthoranalysis}. We attempt to estimate the number of unique authors for each year (blue) as well as the number of new authors per year (orange). As the same astronomer may appear under different name strings on publications of a given year, e.g.~once with full first name, once with abbreviated first name, once with initial, once without, and as the author may or may not have sometimes a single, sometimes a double last name, etc., we sanitise the names into the most plausible combination of last name and abbreviated first name and use these author names to compute the values for the figure. This method necessarily reduces the number of unique authors as e.g. Smith, John and Smith, Jonathan will receive the same abbreviation and be counted as one author. Nevertheless, as we do have author names with full names from some journals, we can estimate the frequency of such conflicts and find that, on average, there are 1.07 different full names for each abbreviated name. For our purpose, this is sufficient: We find that in 2024 4,055 unique authors appeared on ALMA publications, of whom 966 authors had not been on any ALMA publication ever before. This number exceeds the number of newly registered ALMA users per year. On average, over the last 5 years, 881 unique authors appeared for the first time each year on ALMA publications - approximately 20\% more than new authors on proposals - and the total number of unique authors is 9,415.

The increasing number of unique authors is a reflection of the fact that both the total number of ALMA publications (see Figure \ref{figure:evolutiongeneral}) as well as the number of authors on a single publication (see Subsection~\ref{subsection:authors}) are growing with time. 

We can apply the same methodology to the authors on ALMA programs and find that about half of all authors of ALMA publications have never co-authored an ALMA proposal that was accepted and had data delivered.

\begin{figure}[t]
  \resizebox{\hsize}{!}{\includegraphics{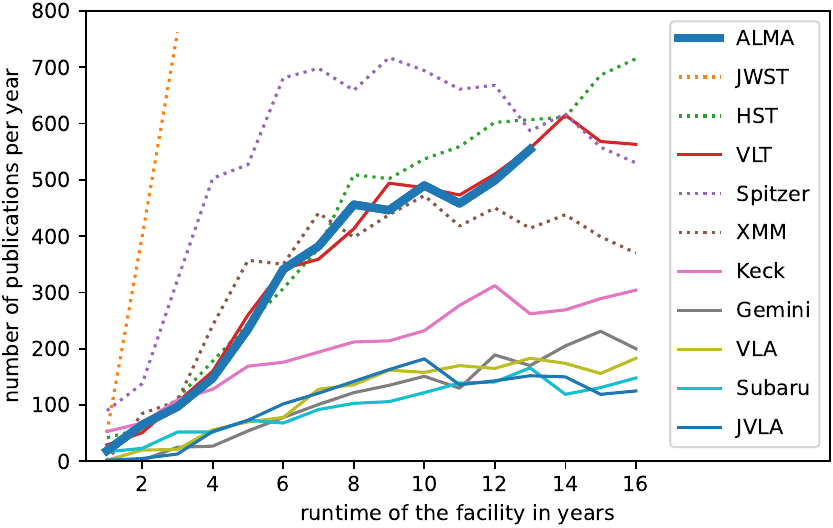}}
  \caption{Evolution of the yearly number of publications for major astronomical facilities as a function of the runtime of the facility. Space based observatories are shown with dotted lines and ground based facilities with solid lines. ALMA's evolution is shown with a thick blue line and is compatible with the evolutions of HST and the VLT in their first years of operations.}
  \label{figure:comparisonwithotherfacilities}
\end{figure}

\subsection{Comparison with other facilities}
\label{section:comparisonwithotherfacilities}
This work concentrates on publication statistics of ALMA. Other authors \citep[e.g.][]{2019AAS...23345302C_Crabtree} tackle comparisons between different observatories. Nevertheless, we include here two figures comparing different astronomical facilities. 

Figure~\ref{figure:comparisonwithotherfacilities}, shows the number of publications per year as a function of the facility's runtime\footnote{Here defined as starting when the first publication using data from the facility appeared.}. Space facilities are shown with dotted lines, and ground-based facilities are shown with solid lines. ALMA's evolution is plotted with a thick blue line. There is a clear distinction visible between space facilities with high publication rates and ground-based facilities with lower publication rates, with the exception of ALMA and the VLT which both track HST's evolution rather well and have tracked and then surpassed XMM {\it Newton}'s evolution, reaching nearly {\it Spitzer}'s 13-year value.

In Section~\ref{subsection:publicationfraction}, we have found that, considering only data for which three years have passed since the data release to the PIs, an average of 69\% of the observed projects have been published through PI research or archival research. This value is comparable to - but slightly lower than - the fraction reported for XMM-Newton (76\%, \citet{2014AN....335..210N_Ness}), using the same cut-off and also excluding ALMACAL-type publications. We can also compare ALMA's publication fraction value to the VLT: \cite{2015Msngr.162....2_Sterzik} report a publication fraction of 44\% without applying any cut-off. The equivalent value for ALMA is 54\%. Comparing publication fraction values of difference facilities is delicate, as these values continue to increase over time (see Section~\ref{subsection:publicationfraction}), which is consistent with the findings from \citet{2025AN....34670014_Ness} showing a publication fraction increase of at least up to ten years after data delivery. For the comparisons with XMM and the VLT here, however, we are fortunate, as the corresponding articles were written at approximately the same telescope runtimes: For the VLT after 16 years and for XMM after 13 years matching the 13 years for this work rather well.

ALMA's scientific output, in terms of the number of publications, is thus similar to that of some of the most prolific space observatories and the VLT. 

As a caveat, we do note, however, that historical publication numbers are not necessarily directly comparable to those of today, as their total number has substantially increased over time, both due to the increasing number of active astronomers, as well as to the increase of the available amount of data. In the first year of HST's operations in 1990, ADS lists about 17 thousand refereed publications for all of astronomy; in the first year of VLT operations, the value has grown to 21 thousand publications, and in the first year of ALMA operations, 28 thousand refereed publications were published per year\footnote{For an order of magnitude estimate of the possible effect, we can normalise the yearly number of publications each facility has by a factor of the total number of refereed astronomy publications of 2024 divided by the total number of such publications when the facility had the same runtime as ALMA has now. We find that HST, {\it Spitzer} and the VLT would have had substantially higher numbers of publications than ALMA, whereas all other facilities listed in Fig.~\ref{figure:comparisonwithotherfacilities} would still have a smaller number of publications per year than ALMA has. No such calculation for JWST can be carried out due to the short operational period.}.

Moreover, an increasing amount of science is multi-wavelength, further elevating the number of publications that make use of ALMA data compared to what was typical in astronomy two or three decades ago. This is also reflected in the evolution of the average number of references in refereed astronomy publications, as can be extracted from ADS. While in 1990 each publication on average referenced 14.5 (median 9) other refereed publications, in 1999, the first year of VLT operations, the value had grown to 20.8 (median 15), and to 35.6 (median 29) by the start of ALMA operations. While the evolution of the number of publications in astronomy is approximately linear over time, the evolution of the number of references in each publication increases at a rate faster than linear. In 2024, the average astronomy publication referenced 60.8 (median 53) other refereed publications.

\begin{figure}[t]
  \resizebox{\hsize}{!}{\includegraphics{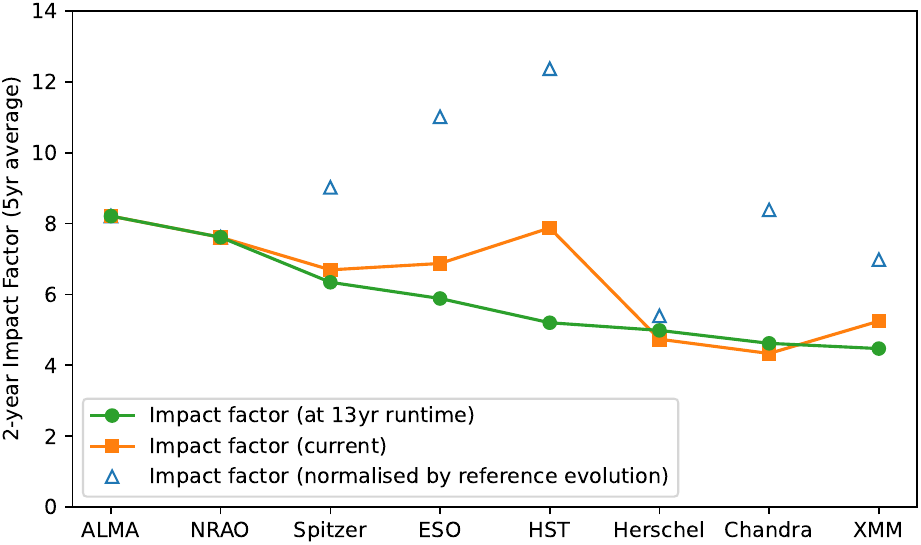}}
  \caption{Two-year impact factors of some of the major facilities that provide data-collections or {\it Bib Groups} in ADS. In each case an average value over five year computed once for the impact factor at the same runtime as ALMA (green) and once at the present day (orange). Lines are plotted to easier guide the eye. Blue triangles show 13-year runtime values scaled by the average number of references in astronomy publications at that particular year compared to 2024.}
  \label{figure:impactfactorcomparison}
\end{figure}

Other caveats apply as well. It is for example plausible to assume that HST's publication rate would have been significantly larger without the primary mirror flaw \citep{NASA1990-HSTFailure} and the time it took to repair it - a glimpse of which can be seen by the steeply rising publication evolution of JWST. The publication evolution of the {\it Spitzer} space telescope is influenced by the fact that once the cryogen was depleted after 5.5 years, as planned, the facility switched to warm mode operations. Data-analysis funding which is available for US authors of publications using some facilities (e.g. {\it Chandra}, XMM, HST, JWST, ALMA) also can skew statistics depending on the amount available as well as compared to facilities (e.g. VLT) and regions that do not have access to such funding. Moreover, another very well-known caveat is that different types of facilities serve different types of communities and different science topics, which also may lead to strong inter-facility differences. 

In Figure~\ref{figure:impactfactorcomparison}, we compare ALMA's impact factor with the impact factors of other major astronomical institutions and facilities. The impact factors are computed using the programmatic access to ADS and the {\it Bib Group} tags. To obtain robust measurements, for all institutes and facilities we average the impact factors of five years (see Subsection~\ref{subsection:impactfactor}). Green markers show the averaged impact factors at the same runtime as ALMA. Orange markers show the current impact factors, i.e. averaged from 2020 to 2024. For the NRAO and ESO {\it Bib Groups}, we have carefully removed the ALMA publications to avoid double counting. 

We find that ALMA's impact factor (identical for both measurements by construction) is larger than the impact factors of all other facilities listed. The same is true for all of the PI-observatory missions of ESA (see Figure~30 of \citet{2024arXiv240212818D_DeMarchi})\footnote{Although the current five-year average impact factor of INTEGRAL is higher, it is not sustained over the course of the mission.}

The calculation of the impact factor - i.e. the average number of citations publications receive after one or two years - is robust against slight differences facilities may have  including or excluding publications in their curated bibliography. In principle, the impact factor calculation is also robust to the increased number of publications over time. However, as we show above, the number of references in each publication is strongly increasing over time, which thus does affect the comparison of impact factor measurements made at different times. To give an order of magnitude indication of this effect, we have scaled the impact factors at fixed runtime (green markers) in Figure~\ref{figure:impactfactorcomparison} by the ratio of the average number of references at the mid-points of the five year averaging intervals leading to the blue markers. After this scaling, ALMA's 13-year runtime impact factor is smaller than those of {\it Spitzer}, ESO, HST and {\it Chandra}.

\section{Summary and Conclusions}
We have analysed the statistics of 4,190 refereed ALMA publications from 2012 to 2024. In this Section, we summarise the results and provide conclusions.

\label{section:conclusions}
\begin{itemize}
    \item The number of ALMA publications per year continues to rise. In 2024, for the first time, more than 500 refereed articles were published. Since 2019, the growth is driven by publications that make, at least partly, use of archival data.
    \item The scientific categories `ISM and star formation', `Active Galaxies', `Galaxy evolution' and `Disk and planet formation' dominate the absolute number of publications. In terms of publications per invested hour of observing time, `Sun' stands out, and in terms of citations per invested hour, the two categories `Disk and planet formation', `Active galaxies' yield the largest values.
    \item Not all data of all projects listed in a publication are actually used. While in nearly all cases the cited Band 6 and Band 7 data were really used in the publications (on average over 90\% of the observing hours), for Band 10, on average only 59\% of the observed hours were actually used. The latter values are however affected by low number statistics.
    \item Similarly, we find that for the scientific productivity measured in hours as well as in citations per invested hour of observing time Band 6 clearly dominates. The lowest values has Band 9 with less than 40\% of the Band 6 productivity for publications and citations, respectively, again affected by low number statistics.
    \item The vast majority (88\%) of all ALMA results are published in just four journals: ApJ (35\%), A\&A (25\%), MNRAS (17\%) and ApJL (11\%). While the fraction of publications in the high-impact journals {\it Nature} and {\it Science} has by now dropped from the maximum value of nearly 9\% in 2013 to roughly 1\% to 2\%, the absolute number on average is 6.1 per year with no sign of decline.
    \item First authors from institutes in 52 countries and six major international organisations (ALMA, ESO, ESA, SKA, IRAM, EHT) have been publishing ALMA data. This is more than double the number of countries that are affiliated with the three ALMA partners. The three most publishing countries in 2024 are USA, Japan and Germany. 
    \item In 2024, China is the fourth-most publishing country despite it not being affiliated with ALMA. About 82\% of all Chinese first-author publications make use of PI data, indicating strong collaboration with research groups from ALMA countries.
    \item We find for all five ALMA regions (CHILE, EA, EU, NA and OTHER) that if the size of the community is larger (smaller) than the time-share of that region, the the fraction of publications from that region is also larger (smaller). I.e. ‘data do not write papers, people write papers’.
    \item The number of authors on publications has been growing from 6 and 9.9 to 10 and 18.9 for the median and mean, respectively. On average, about 80\% of all publications are multi-ALMA-region publications.
    \item On average, each published PI project has been used 3.7 times in different publications. 60\% of the published projects have been used more than once. One PI project has been used 81 times.
    \item In 2024, $\sim$40\% of all ALMA publications were making use of at least some archival PI data. About half of those only used archival PI data. 
    \item 14\% of the published projects have never been published by the PI group but have been published by archival researchers. 
    \item 13\% of all ALMA publications in {\it Nature} and {\it Science} make use of at least some ALMA archival data. 6.4\% make only use of archival ALMA data.
    \item The availability of archival data enables astronomers in countries like India, Vietnam, Brazil, Argentina, and Mexico with less than 15 kUS\$ of GDP per capita, to engage very strongly in archival research, broadening the community beyond the ALMA affiliated countries. 
    \item It takes the PI and collaborators a median of 2.1 years (mean 2.55 years) until the first publication appears and a median of 4.2 years (mean 4.41 years) for any publication they write. This time is significantly longer than the proprietary period of 12 months. The time-to-publication has also increased significantly from the 1.4 years of ALMA’s Cycle 0 projects. This evolution should be monitored closely and mitigated if necessary.
    \item The distribution of time-intervals to the first publication by the PI group can be well fitted by a log-normal distribution, i.e. a Gaussian distribution but with a logarithmic time-axis. The same is true for the proposal submission before the deadline \citep{2017msngr.169...53_Stoehr}.
    \item We find no correlation of the average time it takes the PI group to publish their data with the data size. We do find a slight trend for the median time where projects in the range of one GB to 10 GB of raw data size get typically published roughly 3.6 months faster than projects between one TB and 10 TB. 
    \item Data remain interesting for a very long time. Even in 2024, data from 14 projects of Cycle 0 were used by the PI groups. The longest delay recorded between data delivery and the first publication by the PI group using those data is 10.5 years.
    \item Currently, data from the very recent cycles are far less used in recent publications that data from Cycles 3 to 7. The most popular Cycle used in publications of 2023 and 2024 is still Cycle 3 (project codes 2015.1). 
    \item On average, considering data observed between 2011 up to and including 2021 as well as publications up to and including 2024, we find a publication fraction for ALMA of $\sim$60\% for data published by the PI group, and $\sim$70\% for data published by PI groups as well as by archival researchers.
    \item Both, the fraction of data that get published by the PI group continues to decrease and the time it takes to publish the results from ALMA projects continues to increase. After three years 80\% of the Cycle 0 data (project codes 2011.0) were published by the PI group, whereas only 30\% of the Cycle 7 data (project codes 2019.1) have been published after the same time-span.  Less than 3\% of all projects of Cycle 9 (project cods 2022.1) were published within the 12 month proprietary period.
    \item 22\% of PIs filled out the continuous anonymous questionnaire in case no publication was identified two to three years after data delivery. ‘A publication is in preparation’, clearly dominates with 38\% of the answers. Six percent of the respondents indicates that the expected science was not contained in the data, which is three times lower than the value for the VLT.
    \item Improvements to science operations have the potential to significantly increase the publication fraction, as a total of 27\% of the answers indicate problems with the quality or completeness of the data, no time being available any more or `other' reasons.
    \item ALMA publications have received a total of 169,985 citations. The number of citations per year is rising steeply with time, significantly faster than the number of publications per year. 
    \item Part of this effect is likely related to the increasing number of references authors place into their publications. In 1990 each publication on average referenced 14.4 other refereed publications, in 2024 this value has grown to 60.6.   
    \item On average, a publication receives 40.6 citations. Publications making use of DDT data receive more citations initially as well as more citations overall compared to all ALMA publications.
    \item The impact factor - the average number of citations a one to two year old publication has received - remains remarkably stable at an average of 8.21.
    \item A total of 193 publications made use of SV data with data from Orion KL being used in 34 publications. The first publication appears about five times faster after the data-delivery than for PI data. The use of SV data has dropped from 63\% in 2012 to about 1\% in 2024 which is a natural consequence of the decreasing amount of new observing modes and thus data releases. 
    \item The total number of publications and citations are dominated by Standard projects which dominate the total observing time. Normalised by the invested observing time and averaged, the number of citations of VLBI projects (only normalised by the ALMA observing time, though) exceeds that of DDT programs which in turn exceeds the numbers of LP, ToO and Standard projects which all three have nearly identical numbers and evolution. We note that we find that LPs do have a somewhat larger publication fraction (Figure~\ref{figure:citationsperhourvsarraytimeperprogram}) and maybe a slightly larger publication delay (top-right panel of Figure~\ref{figure:observingmode}).
    \item Normalising the citations by the invested observing time we find no trend with the runtime of the corresponding programs. LP or shorter programs generate similar amounts of citations. When only considering projects that have actually been published, there is a trend for shorter projects resulting typically in slightly more citations than for larger projects.
    \item At the end of 2024, 12,073 users were registered in the ALMA Science Portal. A linear regression fit excluding the initial registration for Cycle 0 yields a slope of 740 new users registering each year.
    \item In 2024, there were 10,498 authors on ALMA publications with a total of 4,058 unique authors and 966 unique authors who had not appeared on an ALMA publication before. Together with the evolution of the registered users in the ALMA Science Portal, we see this as direct evidence for the fulfillment of ALMA's promise to be a facility usable not only by radio interferometry experts but by all astronomers.
    \item About half of all authors of all ALMA publications have never co-authored an observed ALMA proposal.
    \item The evolution of the number of ALMA publications is similar to those of the HST and VLT at the same facility runtime. 
    \item ALMA's two-year impact factor is larger than all other major PI-observation facilities at the same runtime, although it is important to recognize that over time the number of references per publication increases, which biases this finding towards more recent facilities. It can be assumed that the five year value for JWST will be larger than that of ALMA once it can be computed.
    \item The list of the most cited publications is dominated by the EHT project to which ALMA's observations made pivotal contributions.
\end{itemize}

ALMA has been able to attract a large user community, which increases every year. More than 4,000 unique authors appeared on ALMA publications in 2024 alone, nearly 1,000 of whom had never authored an ALMA publication before. We attribute this not only to ALMA’s technical capabilities, but also to the large investment of the observatory in providing high-level data products, in the ease of data discovery and data access, and in the extended user support. The strength of the community is also reflected by the fact that 76\% of the published data have been used in more than one publication, as well as by the high popularity, even in 2024, of data observed nearly a decade ago (e.g. Cycle 3).

Consistent with this large and now truly global community, the number of publications continues to increase. Importantly, since 2019, the increased number of publications is due to the use of archival data, purely or together with PI data, indicating the high and long-term value of ALMA data and the importance of the ALMA Science Archive. Remarkably, half of the authors on ALMA publications have never been co-author of an observed ALMA proposal. 

However, we note that for ALMA, the fraction of projects that have been published a given number of years after data delivery continues to decrease over time. This trend should be monitored closely, its causes investigated, and appropriate measures adopted to ensure that the effort and investment devoted to carrying out observations continues to yield scientific value at the highest level. 
 
ALMA is a high-impact observatory but we highlight that given the observed publication delay, the impact reported here reflects mostly the impact of publications up to Cycle 7. Given the evolution of astronomical research towards an increased weight set to time-domain and multi-messenger observations, we expect that the ability of ALMA to build synergies with other observatories and facilitate such synergetic, often time-sensitive, observations will be crucial to keep and increase the already excellent productivity and impact in the coming years. 

Consistent with the high impact factor, an average of 6.1 papers are published per year in the high impact journals {\it Nature} and {\it Science} based on ALMA data, and iconic papers like the first image of a black hole shadow \citep{2019ApJ...875L...1_EHT} or the high resolution image of the protoplanetary disc around HL Tau \citep{2015ApJ...808L...3A_ALMA} have already gathered more than 3,000 and 1,000 citations, respectively.

Our findings in this work highlight ALMA’s substantial productivity and the wide-reaching impact of its scientific contributions.

\clearpage
 
\section*{Acknowledgments}
We are grateful for discussions with Ignacio Toledo, Suzanna Randall, Andrea Corvill\'on, John Carpenter, Daisuke Iono, Sean Dougherty, Xavier Barcons, Michael Sterzik, Martino Romaniello and Nando Patat. ALMA is a partnership of ESO (representing its member states), NSF (USA) and NINS (Japan), together with NRC (Canada), NSTC and ASIAA (Taiwan), and KASI (Republic of Korea), in cooperation with the Republic of Chile. The Joint ALMA Observatory is operated by ESO, AUI/NRAO and NAOJ. This research has made use of the Astrophysics Data System, funded by NASA under Cooperative Agreement 80NSSC21M00561.

\vspace{0.25cm}

\bibliographystyle{aa}
\bibliography{almapublicationstatistics}

\appendix
\label{appendix}
\addcontentsline{toc}{section}{Appendix}

\begin{table*}[!htbp]
\caption{The 50 ALMA publications between 2012 and 2023 with the largest average citation rate. Publications from 2024 have been skipped to avoid short-term effects.}
\label{table:largestcitationrate}
\centering
\begin{tabular}{r r l l}
\hline\hline
\input{table_04.tex}
\end{tabular}
\end{table*}

\begin{table*}[h]
\caption{The 50 ALMA projects that were used in the largest number of publications between 2012 and 2024. Publications that make use of large fractions of the ALMA archive (e.g. ALMACAL) have been excluded from the counting.}
\label{table:mostoseddatasets}
\centering
\begin{tabular}{r l l}
\hline\hline
\input{table_05.tex}
\end{tabular}
\end{table*}

\end{document}

%% file: table_01.tex
    Science category               & observed & published & publ. & citations & citations & citations & citation &   publ. & citations\\
                                & projects &  projects &   nr. &           & / project &   / publ. & fraction & / hour  & / hour \\
\hline
Cosmology                      &      284 &       162 &   216 &      7211 &     44.4 &     33.5 &     4.5 \% &    0.09 &     3.2 \\
Galaxy evolution               &      906 &       543 &   700 &     31780 &     58.5 &     45.4 &    20.0 \% &    0.08 &     3.7 \\
Active galaxies                &     1036 &       568 &   721 &     38827 &     68.4 &     53.8 &    24.5 \% &    0.09 &     4.9 \\
Local Universe                 &      141 &        68 &   111 &      3905 &     57.9 &     35.2 &     2.5 \% &    0.10 &     3.5 \\
ISM and star formation         &     1338 &       604 &  1162 &     34566 &     57.3 &     29.7 &    21.8 \% &    0.12 &     3.5 \\
Stars and stellar evolution    &      348 &       190 &   250 &      5694 &     29.9 &     22.8 &     3.6 \% &    0.13 &     3.0 \\
Disks and planet formation     &      704 &       430 &   736 &     34285 &     79.8 &     46.6 &    21.6 \% &    0.15 &     6.8 \\
Solar system                   &      135 &        50 &    69 &      1764 &     35.6 &     25.7 &     1.1 \% &    0.11 &     2.8 \\
Sun                            &       37 &        25 &    41 &       618 &     24.7 &     15.2 &     0.4 \% &    0.25 &     3.8 \\
&  &  &  &  & \\ 
Total (PI projects)            &     4929 &     2639 &    4005 &   158650 &$\varnothing$ 60.1 &$\varnothing$  39.6 &     100 \% &$\varnothing$  0.11 &$\varnothing$   4.3 \\
\hline
&  &  &  &  & \\ 
Total (all ALMA projects)            &     4962 &     2670 &    4190 &   169985 &$\varnothing$ 63.7 &$\varnothing$  40.6 &     100 \% &$\varnothing$  0.11 &$\varnothing$   4.5 \\

%% file: table_02.tex
Band & publ. &  fraction & fraction &     citation & citations \\
        &  number   &         &  norm. &  fraction & / hour \\
\hline
  3 &  1259 &         30.0 \% &         20.1 \% &          15.2 \% &        2.5 \\ 
  4 &   336 &          8.0 \% &          5.4 \% &           1.7 \% &        1.3 \\ 
  5 &   143 &          3.4 \% &          2.3 \% &           0.5 \% &        0.9 \\ 
  6 &  2469 &         58.9 \% &         39.4 \% &          51.3 \% &        6.4 \\ 
  7 &  1643 &         39.2 \% &         26.2 \% &          27.9 \% &        5.7 \\ 
  8 &   213 &          5.1 \% &          3.4 \% &           1.1 \% &        1.7 \\ 
  9 &   189 &          4.5 \% &          3.0 \% &           2.3 \% &        5.8 \\ 
 10 &    11 &          0.3 \% &          0.2 \% &           0.1 \% &        0.8 \\ 
\hline 

%% file: table_03.tex
    Journal & fraction of & publi- &  citations & citations / \\
            & publications & cations &  & publication\\
    \hline
    ApJ            &     35\%    &     1464    &     58601   & 40.0 \\ 
A\&A           &     25\%    &     1041    &     32746   & 31.5 \\ 
MNRAS          &     17\%    &      712    &     19604   & 27.5 \\ 
ApJL           &     11\%    &      444    &     35859   & 80.8 \\ 
AJ             &      2\%    &       95    &      2133   & 22.5 \\ 
PASJ           &      2\%    &       84    &      2009   & 23.9 \\ 
ApJS           &      1\%    &       60    &      3005   & 50.1 \\ 
Nature         &      1\%    &       58    &      7942   & 136.9 \\ 
NatAs          &      1\%    &       53    &      2787   & 52.6 \\ 
Science        &      1\%    &       21    &      2516   & 119.8 \\ 
{\it other}    &      4\%    &      158    &      2783   &  17.6 \\ 
\hline 

%% file: table_04.tex
citations/yr & citations & first author: publication title & BibCode \\ 
\hline 
613   &   3375  & EHT Collaboration: First M87 Event Horizon Telescope Results. I. The Shadow & \href{https://almascience.org/aq/?bibCode=2019ApJ...875L...1E&result_view=publications}{2019ApJ...875L...1E} \\ 
458   &   1145  & EHT Collaboration: First Sagittarius A* Event Horizon Telescope Results. I. & \href{https://almascience.org/aq/?bibCode=2022ApJ...930L..12E&result_view=publications}{2022ApJ...930L..12E} \\ 
239   &   1319  & EHT Collaboration: First M87 Event Horizon Telescope Results. VI. The Shado & \href{https://almascience.org/aq/?bibCode=2019ApJ...875L...6E&result_view=publications}{2019ApJ...875L...6E} \\ 
224   &   1233  & EHT Collaboration: First M87 Event Horizon Telescope Results. V. Physical O & \href{https://almascience.org/aq/?bibCode=2019ApJ...875L...5E&result_view=publications}{2019ApJ...875L...5E} \\ 
222   &   1221  & EHT Collaboration: First M87 Event Horizon Telescope Results. IV. Imaging t & \href{https://almascience.org/aq/?bibCode=2019ApJ...875L...4E&result_view=publications}{2019ApJ...875L...4E} \\ 
186   &   465  & EHT Collaboration: First Sagittarius A* Event Horizon Telescope Results. VI & \href{https://almascience.org/aq/?bibCode=2022ApJ...930L..17E&result_view=publications}{2022ApJ...930L..17E} \\ 
159   &   876  & EHT Collaboration: First M87 Event Horizon Telescope Results. II. Array and & \href{https://almascience.org/aq/?bibCode=2019ApJ...875L...2E&result_view=publications}{2019ApJ...875L...2E} \\ 
155   &   389  & EHT Collaboration: First Sagittarius A* Event Horizon Telescope Results. V. & \href{https://almascience.org/aq/?bibCode=2022ApJ...930L..16E&result_view=publications}{2022ApJ...930L..16E} \\ 
146   &   512  & EHT Collaboration: First M87 Event Horizon Telescope Results. VIII. Magneti & \href{https://almascience.org/aq/?bibCode=2021ApJ...910L..13E&result_view=publications}{2021ApJ...910L..13E} \\ 
146   &   951  & Andrews, Sean M.: The Disk Substructures at High Angular Resolution Projec & \href{https://almascience.org/aq/?bibCode=2018ApJ...869L..41A&result_view=publications}{2018ApJ...869L..41A} \\ 
138   &   763  & EHT Collaboration: First M87 Event Horizon Telescope Results. III. Data Pro & \href{https://almascience.org/aq/?bibCode=2019ApJ...875L...3E&result_view=publications}{2019ApJ...875L...3E} \\ 
134   &   1278  & ALMA Partnership: The 2014 ALMA Long Baseline Campaign: First Results from & \href{https://almascience.org/aq/?bibCode=2015ApJ...808L...3A&result_view=publications}{2015ApJ...808L...3A} \\ 
125   &   313  & EHT Collaboration: First Sagittarius A* Event Horizon Telescope Results. II & \href{https://almascience.org/aq/?bibCode=2022ApJ...930L..14E&result_view=publications}{2022ApJ...930L..14E} \\ 
109   &   384  & Wang, Feige: A Luminous Quasar at Redshift 7.642 & \href{https://almascience.org/aq/?bibCode=2021ApJ...907L...1W&result_view=publications}{2021ApJ...907L...1W} \\ 
108   &   270  & EHT Collaboration: First Sagittarius A* Event Horizon Telescope Results. IV & \href{https://almascience.org/aq/?bibCode=2022ApJ...930L..15E&result_view=publications}{2022ApJ...930L..15E} \\ 
106   &   267  & EHT Collaboration: First Sagittarius A* Event Horizon Telescope Results. II & \href{https://almascience.org/aq/?bibCode=2022ApJ...930L..13E&result_view=publications}{2022ApJ...930L..13E} \\ 
99   &   349  & EHT Collaboration: First M87 Event Horizon Telescope Results. VII. Polariza & \href{https://almascience.org/aq/?bibCode=2021ApJ...910L..12E&result_view=publications}{2021ApJ...910L..12E} \\ 
97   &   733  & Chatterjee, S.: A direct localization of a fast radio burst and its host & \href{https://almascience.org/aq/?bibCode=2017Natur.541...58C&result_view=publications}{2017Natur.541...58C} \\ 
93   &   607  & Tacconi, L. J.: PHIBSS: Unified Scaling Relations of Gas Depletion Time  & \href{https://almascience.org/aq/?bibCode=2018ApJ...853..179T&result_view=publications}{2018ApJ...853..179T} \\ 
90   &   315  & Leroy, Adam K.: PHANGS-ALMA: Arcsecond CO(2-1) Imaging of Nearby Star-fo & \href{https://almascience.org/aq/?bibCode=2021ApJS..257...43L&result_view=publications}{2021ApJS..257...43L} \\ 
75   &   188  & Bouwens, R. J.: Reionization Era Bright Emission Line Survey: Selection  & \href{https://almascience.org/aq/?bibCode=2022ApJ...931..160B&result_view=publications}{2022ApJ...931..160B} \\ 
70   &   247  & Kocherlakota, Prashant: Constraints on black-hole charges with the 2017 EHT obse & \href{https://almascience.org/aq/?bibCode=2021PhRvD.103j4047K&result_view=publications}{2021PhRvD.103j4047K} \\ 
68   &   444  & Long, Feng: Gaps and Rings in an ALMA Survey of Disks in the Taurus  & \href{https://almascience.org/aq/?bibCode=2018ApJ...869...17L&result_view=publications}{2018ApJ...869...17L} \\ 
68   &   583  & Ansdell, M.: ALMA Survey of Lupus Protoplanetary Disks. I. Dust and G & \href{https://almascience.org/aq/?bibCode=2016ApJ...828...46A&result_view=publications}{2016ApJ...828...46A} \\ 
67   &   302  & Yang, Jinyi: Pniu'ena: A Luminous z = 7.5 Quasar Hosting a 1.5 Bill & \href{https://almascience.org/aq/?bibCode=2020ApJ...897L..14Y&result_view=publications}{2020ApJ...897L..14Y} \\ 
65   &   427  & Huang, Jane: The Disk Substructures at High Angular Resolution Projec & \href{https://almascience.org/aq/?bibCode=2018ApJ...869L..42H&result_view=publications}{2018ApJ...869L..42H} \\ 
62   &   93  & Lee, Janice C.: The PHANGS-JWST Treasury Survey: Star Formation, Feedbac & \href{https://almascience.org/aq/?bibCode=2023ApJ...944L..17L&result_view=publications}{2023ApJ...944L..17L} \\ 
61   &   92  & Furtak, Lukas J.: JWST UNCOVER: Extremely Red and Compact Object at z <SUB & \href{https://almascience.org/aq/?bibCode=2023ApJ...952..142F&result_view=publications}{2023ApJ...952..142F} \\ 
60   &   91  & Lu, Ru-Sen: A ring-like accretion structure in M87 connecting its bl & \href{https://almascience.org/aq/?bibCode=2023Natur.616..686L&result_view=publications}{2023Natur.616..686L} \\ 
58   &   204  & Öberg, Karin I.: Molecules with ALMA at Planet-forming Scales (MAPS). I.  & \href{https://almascience.org/aq/?bibCode=2021ApJS..257....1O&result_view=publications}{2021ApJS..257....1O} \\ 
57   &   86  & Ohashi, Nagayoshi: Early Planet Formation in Embedded Disks (eDisk). I. Ove & \href{https://almascience.org/aq/?bibCode=2023ApJ...951....8O&result_view=publications}{2023ApJ...951....8O} \\ 
56   &   476  & Andrews, Sean M.: Ringed Substructure and a Gap at 1 au in the Nearest Pro & \href{https://almascience.org/aq/?bibCode=2016ApJ...820L..40A&result_view=publications}{2016ApJ...820L..40A} \\ 
55   &   248  & Chevance, Mélanie: The lifecycle of molecular clouds in nearby star-forming & \href{https://almascience.org/aq/?bibCode=2020MNRAS.493.2872C&result_view=publications}{2020MNRAS.493.2872C} \\ 
55   &   83  & Algera, Hiddo S. B.: The ALMA REBELS survey: the dust-obscured cosmic star fo & \href{https://almascience.org/aq/?bibCode=2023MNRAS.518.6142A&result_view=publications}{2023MNRAS.518.6142A} \\ 
54   &   191  & Benisty, Myriam: A Circumplanetary Disk around PDS70c & \href{https://almascience.org/aq/?bibCode=2021ApJ...916L...2B&result_view=publications}{2021ApJ...916L...2B} \\ 
54   &   243  & Tobin, John J.: The VLA/ALMA Nascent Disk and Multiplicity (VANDAM) Surv & \href{https://almascience.org/aq/?bibCode=2020ApJ...890..130T&result_view=publications}{2020ApJ...890..130T} \\ 
54   &   355  & Zhang, Shangjia: The Disk Substructures at High Angular Resolution Projec & \href{https://almascience.org/aq/?bibCode=2018ApJ...869L..47Z&result_view=publications}{2018ApJ...869L..47Z} \\ 
54   &   81  & Furtak, Lukas J.: UNCOVERing the extended strong lensing structures of Abe & \href{https://almascience.org/aq/?bibCode=2023MNRAS.523.4568F&result_view=publications}{2023MNRAS.523.4568F} \\ 
52   &   338  & Ansdell, M.: ALMA Survey of Lupus Protoplanetary Disks. II. Gas Disk  & \href{https://almascience.org/aq/?bibCode=2018ApJ...859...21A&result_view=publications}{2018ApJ...859...21A} \\ 
51   &   128  & Currie, Thayne: Images of embedded Jovian planet formation at a wide sep & \href{https://almascience.org/aq/?bibCode=2022NatAs...6..751C&result_view=publications}{2022NatAs...6..751C} \\ 
51   &   439  & Scoville, N.: ISM Masses and the Star formation Law at Z = 1 to 6: ALM & \href{https://almascience.org/aq/?bibCode=2016ApJ...820...83S&result_view=publications}{2016ApJ...820...83S} \\ 
50   &   228  & Dudzeviit, U.: An ALMA survey of the SCUBA-2 CLS UDS field: physical pr & \href{https://almascience.org/aq/?bibCode=2020MNRAS.494.3828D&result_view=publications}{2020MNRAS.494.3828D} \\ 
50   &   326  & Dullemond, Cornelis P.: The Disk Substructures at High Angular Resolution Projec & \href{https://almascience.org/aq/?bibCode=2018ApJ...869L..46D&result_view=publications}{2018ApJ...869L..46D} \\ 
49   &   173  & Greaves, Jane S.: Phosphine gas in the cloud decks of Venus & \href{https://almascience.org/aq/?bibCode=2021NatAs...5..655G&result_view=publications}{2021NatAs...5..655G} \\ 
49   &   272  & Fluetsch, A.: Cold molecular outflows in the local Universe and their  & \href{https://almascience.org/aq/?bibCode=2019MNRAS.483.4586F&result_view=publications}{2019MNRAS.483.4586F} \\ 
48   &   219  & Béthermin, M.: The ALPINE-ALMA [CII] survey: Data processing, catalogs, & \href{https://almascience.org/aq/?bibCode=2020A&ampA...643A...2B&result_view=publications}{2020A\&A...643A...2B} \\ 
48   &   410  & Pascucci, I.: A Steeper than Linear Disk MassStellar Mass Scaling Rel & \href{https://almascience.org/aq/?bibCode=2016ApJ...831..125P&result_view=publications}{2016ApJ...831..125P} \\ 
47   &   311  & Hashimoto, Takuya: The onset of star formation 250 million years after the  & \href{https://almascience.org/aq/?bibCode=2018Natur.557..392H&result_view=publications}{2018Natur.557..392H} \\ 
47   &   354  & Dunlop, J. S.: A deep ALMA image of the Hubble Ultra Deep Field & \href{https://almascience.org/aq/?bibCode=2017MNRAS.466..861D&result_view=publications}{2017MNRAS.466..861D} \\ 
46   &   117  & Chevance, Mélanie: Pre-supernova feedback mechanisms drive the destruction  & \href{https://almascience.org/aq/?bibCode=2022MNRAS.509..272C&result_view=publications}{2022MNRAS.509..272C} \\ 
\hline 

%% file: table_05.tex
publications & Principal Investigator: proposal title & project code \\
    \hline
    81 & Leroy: How Does Cloud-Scale Physics Drive Galaxy Evolution? & \href{https://almascience.org/aq/?projectCode=2015.1.00956.S}{2015.1.00956.S} \\ 
76 & Schinnerer: 100,000 Molecular Clouds Across the Main Sequence: GMCs as the Drive ... & \href{https://almascience.org/aq/?projectCode=2017.1.00886.L}{2017.1.00886.L} \\ 
68 & Schinnerer: The Role of Galactic Environment in GMC and Star Formation & \href{https://almascience.org/aq/?projectCode=2012.1.00650.S}{2012.1.00650.S} \\ 
65 & Blanc: Promoting Diversity: ISM Physics and Star Formation across Different Envi ... & \href{https://almascience.org/aq/?projectCode=2015.1.00925.S}{2015.1.00925.S} \\ 
58 & Sakamoto: From Bars to CMZs and YMCs & \href{https://almascience.org/aq/?projectCode=2013.1.01161.S}{2013.1.01161.S} \\ 
51 & Leroy: Completing a Census of 50pc ISM and Star Formation Properties in Disk Gal ... & \href{https://almascience.org/aq/?projectCode=2018.1.01651.S}{2018.1.01651.S} \\ 
50 & Andrews: Small-Scale Substructures in Protoplanetary Disks & \href{https://almascience.org/aq/?projectCode=2016.1.00484.L}{2016.1.00484.L} \\ 
49 & Jorgensen: Formation of complex organics in solar-type protostars & \href{https://almascience.org/aq/?projectCode=2013.1.00278.S}{2013.1.00278.S} \\ 
46 & Blanc: Completing a Census of Cloud-Scale ISM Structure in Low Mass Disk Galaxie ... & \href{https://almascience.org/aq/?projectCode=2017.1.00392.S}{2017.1.00392.S} \\ 
42 & Le Fèvre: ALPINE: The ALMA Large Program to INvestigate CII at Early times & \href{https://almascience.org/aq/?projectCode=2017.1.00428.L}{2017.1.00428.L} \\ 
42 & Perez: Dust growth in protoplanetary disks: where in the disk are grains growing ... & \href{https://almascience.org/aq/?projectCode=2013.1.00498.S}{2013.1.00498.S} \\ 
42 & EHT Collaboration: Imaging the Black Hole Shadow and Jet Launching Region of M87 & \href{https://almascience.org/aq/?projectCode=2016.1.01154.V}{2016.1.01154.V} \\ 
41 & Oberg: A survey of deuterium chemistry in protoplanetary disks & \href{https://almascience.org/aq/?projectCode=2013.1.00226.S}{2013.1.00226.S} \\ 
39 & Oberg: The Chemistry of Planet Formation & \href{https://almascience.org/aq/?projectCode=2018.1.01055.L}{2018.1.01055.L} \\ 
34 & Liu: On the origin of the dense gas star formation law in Galactic high-mass sta ... & \href{https://almascience.org/aq/?projectCode=2019.1.00685.S}{2019.1.00685.S} \\ 
34 & ALMA observatory: Science verification observation of Orion KL & \href{https://almascience.org/aq/?projectCode=2011.0.00009.SV}{2011.0.00009.SV} \\ 
33 & Sakamoto: Molecular Clouds and Star Formation: Inner Disk of M83 & \href{https://almascience.org/aq/?projectCode=2015.1.00121.S}{2015.1.00121.S} \\ 
33 & Sakamoto: Molecular Clouds and Star Formation: Across M83 & \href{https://almascience.org/aq/?projectCode=2016.1.00386.S}{2016.1.00386.S} \\ 
32 & Johnson: ALMA-LEGUS: The Impact of Spiral Arm Structure on Molecular Cloud Prope ... & \href{https://almascience.org/aq/?projectCode=2015.1.00782.S}{2015.1.00782.S} \\ 
31 & Faesi: Physics at High Angular Resolution in Nearby Galaxies: The Local Galaxy I ... & \href{https://almascience.org/aq/?projectCode=2018.1.01321.S}{2018.1.01321.S} \\ 
30 & ALMA observatory: Science verification observation of HL Tau & \href{https://almascience.org/aq/?projectCode=2011.0.00015.SV}{2011.0.00015.SV} \\ 
29 & Espada: Probing the Embedded Disk of the Giant Elliptical NGC 5128 (Centaurus A) & \href{https://almascience.org/aq/?projectCode=2013.1.00803.S}{2013.1.00803.S} \\ 
29 & Chevance: From the main sequence to the red cloud: linking the molecular cloud l ... & \href{https://almascience.org/aq/?projectCode=2017.1.00766.S}{2017.1.00766.S} \\ 
28 & Scoville: Evolution of ISM, Star Formation and Starbursts & \href{https://almascience.org/aq/?projectCode=2015.1.00137.S}{2015.1.00137.S} \\ 
28 & Faesi: Physics at High Angular Resolution in Nearby Galaxies: The Local Galaxy I ... & \href{https://almascience.org/aq/?projectCode=2018.A.00062.S}{2018.A.00062.S} \\ 
27 & Capak: Measuring the Infrared Emission and Dynamics for a Population of Primordi ... & \href{https://almascience.org/aq/?projectCode=2012.1.00523.S}{2012.1.00523.S} \\ 
26 & Herczeg: An unbiased survey of disk structures in Taurus & \href{https://almascience.org/aq/?projectCode=2016.1.01164.S}{2016.1.01164.S} \\ 
26 & Walter: ASPECS: The ALMA SPECtral line Survey in the UDF - An ALMA Large Program & \href{https://almascience.org/aq/?projectCode=2016.1.00324.L}{2016.1.00324.L} \\ 
25 & Elbaz: Towards a census of star-formation since z~6 with ALMA-1.1mm & \href{https://almascience.org/aq/?projectCode=2015.1.00543.S}{2015.1.00543.S} \\ 
25 & Aravena: Unveiling the population of high-redshift submillimeter galaxies with A ... & \href{https://almascience.org/aq/?projectCode=2013.1.00118.S}{2013.1.00118.S} \\ 
25 & Bouwens: REBELS: An ALMA Large Program to Discover the Most Luminous [CII]+[OIII ... & \href{https://almascience.org/aq/?projectCode=2019.1.01634.L}{2019.1.01634.L} \\ 
24 & Akiyama: Probing disk structure in a cavity of pre-transitional disks around Sun ... & \href{https://almascience.org/aq/?projectCode=2015.1.00888.S}{2015.1.00888.S} \\ 
24 & Scoville: Evolution of ISM in Star-Forming Galaxies at z = 1 - 5 & \href{https://almascience.org/aq/?projectCode=2013.1.00034.S}{2013.1.00034.S} \\ 
24 & ALMA observatory: GRB 110715A followup Band 7 & \href{https://almascience.org/aq/?projectCode=2011.0.00001.CAL}{2011.0.00001.CAL} \\ 
24 & Kohno: ALMA Lensing Cluster Survey & \href{https://almascience.org/aq/?projectCode=2018.1.00035.L}{2018.1.00035.L} \\ 
24 & Belloche: Expanding the frontiers of chemical complexity with ALMA & \href{https://almascience.org/aq/?projectCode=2011.0.00017.S}{2011.0.00017.S} \\ 
23 & Isella: ALMA measurements of disk turbulence & \href{https://almascience.org/aq/?projectCode=2013.1.00601.S}{2013.1.00601.S} \\ 
23 & ALMA observatory: Solar observations & \href{https://almascience.org/aq/?projectCode=2011.0.00020.SV}{2011.0.00020.SV} \\ 
23 & Bauer: Lensing Through Cosmic Time: ALMA Constraints on "Normal" Galaxies in the ... & \href{https://almascience.org/aq/?projectCode=2013.1.00999.S}{2013.1.00999.S} \\ 
22 & Karim: Timing the birth of the red sequence & \href{https://almascience.org/aq/?projectCode=2012.1.00978.S}{2012.1.00978.S} \\ 
22 & Inami: Properties of a temperature-unbiased sample of Herschel 250um-selected ga ... & \href{https://almascience.org/aq/?projectCode=2015.1.01074.S}{2015.1.01074.S} \\ 
21 & Williams: Disk Demographics in Lupus & \href{https://almascience.org/aq/?projectCode=2013.1.00220.S}{2013.1.00220.S} \\ 
21 & Dunlop: An ALMA 1.3-mm image of The Hubble Ultra Deep Field & \href{https://almascience.org/aq/?projectCode=2012.1.00173.S}{2012.1.00173.S} \\ 
21 & Yamamoto: Fifty AU STudy of the chemistry in the disk/envelope system of Solar-l ... & \href{https://almascience.org/aq/?projectCode=2018.1.01205.L}{2018.1.01205.L} \\ 
21 & Miettinen: Size matters: resolving the rest-frame far-infrared-emitting region o ... & \href{https://almascience.org/aq/?projectCode=2016.1.00478.S}{2016.1.00478.S} \\ 
20 & Smail: AS2UDS : Clustering of ~1000 ALMA-identified submillimeter galaxies & \href{https://almascience.org/aq/?projectCode=2015.1.01528.S}{2015.1.01528.S} \\ 
20 & Daddi: A survey for CO[5-4]emission in star forming galaxies at 1.1<z<1.7 & \href{https://almascience.org/aq/?projectCode=2015.1.00260.S}{2015.1.00260.S} \\ 
20 & Lu: A Spectral Line Snapshot Proposal for ALMA: Characterizing Star Formation Ra ... & \href{https://almascience.org/aq/?projectCode=2015.1.00388.S}{2015.1.00388.S} \\ 
20 & Smail: More than LESS: The first fully-identified submillimetre survey & \href{https://almascience.org/aq/?projectCode=2011.0.00294.S}{2011.0.00294.S} \\ 
20 & Belloche: Expanding the frontiers of chemical complexity with ALMA & \href{https://almascience.org/aq/?projectCode=2012.1.00012.S}{2012.1.00012.S} \\ 
\hline 